\documentclass[12pt,letterpaper]{article}

\usepackage{xcolor, colortbl}
\usepackage{amssymb, amsmath, amsthm}
\usepackage{graphicx}
\usepackage{natbib}
\usepackage{float}
\usepackage{setspace}
\usepackage{algorithm}
\usepackage{algorithmic}
\usepackage{xcolor}

\newtheorem{lemma}{Lemma}
\newtheorem{proposition}{Proposition}
\newtheorem{corollary}{Corollary}

\newtheorem{assumption}{Assumption}
\definecolor{lightgray}{gray}{0.9}

\newcommand{\ba}{\mathbf{a}}
\newcommand{\bA}{\mathbf{A}}

\newcommand{\bB}{\mathbf{B}}

\newcommand{\bC}{\mathbf{C}}
\newcommand{\bD}{\mathbf{D}}
\newcommand{\bE}{\mathbf{E}}
\newcommand{\bff}{\mathbf{f}}

\newcommand{\bg}{\mathbf{g}}
\newcommand{\bG}{\mathbf{G}}
\newcommand{\bh}{\mathbf{h}}
\newcommand{\bH}{\mathbf{H}}
\newcommand{\bfI}{\mathbf{I}}
\newcommand{\bJ}{\mathbf{J}}
\newcommand{\bK}{\mathbf{K}}
\newcommand{\bL}{\mathbf{L}}
\newcommand{\bl}{\mathbf{l}}
\newcommand{\bm}{\mathbf{m}}
\newcommand{\bR}{\mathbf{R}}
\newcommand{\bS}{\mathbf{S}}
\newcommand{\bP}{\mathbf{P}}
\newcommand{\bQ}{\mathbf{Q}}
\newcommand{\bx}{\mathbf{x}}
\newcommand{\bX}{\mathbf{X}}
\newcommand{\bu}{\mathbf{u}}

\newcommand{\bY}{\mathbf{Y}}
\newcommand{\bV}{\mathbf{V}}
\newcommand{\bv}{\mathbf{v}}
\newcommand{\bW}{\mathbf{W}}
\newcommand{\bw}{\mathbf{w}}
\newcommand{\by}{\mathbf{y}}
\newcommand{\bz}{\mathbf{z}}
\newcommand{\bZ}{\mathbf{Z}}

\renewcommand{\epsilon}{\varepsilon}
\renewcommand{\hat}{\widehat}
\renewcommand{\tilde}{\widetilde}
\renewcommand{\leq}{\leqslant}
\renewcommand{\geq}{\geqslant}
\newcommand{\argmax}{\mathop{\rm argmax}}
\newcommand{\argmin}{\mathop{\rm argmin}}

\newcommand{\distn}[1]{\mathcal{#1}}
\newcommand{\Em}{\mathbb E}
\newcommand{\Pm}{\mathbb P}

\newcommand{\gvn}{\,|\,}
\newcommand{\di}{\text{d}}
\newcommand{\e}{\text{e}}

\newcommand{\var}{\text{Var}}

\newcommand{\vect}[1]{\boldsymbol #1}

\newcommand{\vbeta}{\vect{\beta}}
\newcommand{\vdelta}{\vect{\delta}}
\newcommand{\vepsilon}{\vect{\epsilon}}

\newcommand{\vGamma}{\vect{\Gamma}}

\newcommand{\vLambda}{\vect{\Lambda}}
\newcommand{\vmu}{\vect{\mu}}
\newcommand{\vOmega}{\vect{\Omega}}
\newcommand{\vphi}{\vect{\phi}}
\newcommand{\vPhi}{\vect{\Phi}}

\newcommand{\vsigma}{\vect{\sigma}}
\newcommand{\vSigma}{\vect{\Sigma}}

\newcommand{\vtheta}{\vect{\theta}}
\newcommand{\vTheta}{\vect{\Theta}}

\newcommand{\bigcdot}{\boldsymbol{\cdot}}
\newcommand{\matlab}{\mathrm{M}\mathrm{{\scriptstyle ATLAB}}}

\DeclareMathOperator{\diag}{diag}
\DeclareMathOperator{\rank}{rank}

\bibpunct{(}{)}{;}{a}{,}{,}

\begin{document}

\title{Large Bayesian VARs with Factor Stochastic Volatility: 
\mbox{Identification, Order Invariance and Structural Analysis}}

\author{Joshua C.C. Chan \\
 {\small Purdue University} \\
  \and Eric Eisenstat \\
 {\small University of Queensland}
  \and Xuewen Yu \\
 {\small Purdue University} \\
}

\date{July 2022}

\maketitle

\onehalfspacing

\begin{abstract} \noindent Vector autoregressions (VARs) with multivariate stochastic volatility are widely used for structural analysis. Often the structural model identified through economically meaningful restrictions---e.g., sign restrictions---is supposed to be independent of how the dependent variables are ordered. But since the reduced-form model is not order invariant, results from the structural analysis depend on the order of the variables. We consider a VAR based on the factor stochastic volatility that is constructed to be order invariant. We show that the presence of multivariate stochastic volatility allows for statistical identification of the model. We further prove that, with a suitable set of sign restrictions, the corresponding structural model is point-identified. An additional appeal of the proposed approach is that it can easily handle a large number of dependent variables as well as sign restrictions. We demonstrate the methodology through a structural analysis in which we use a 20-variable VAR with sign restrictions to identify 5 structural shocks.

\bigskip


\bigskip


\end{abstract}

\thispagestyle{empty}

\newpage

\section{Introduction}

Bayesian vector autoregressions (VARs) with multivariate stochastic volatility, first developed in \citet{CS05} and \citet{primiceri05}, are now the workhorse models in empirical macroeconomics. These multivariate stochastic volatility models, however, have the undesirable property that the implied likelihoods are not invariant to the order of the dependent variables.\footnote{This non-invariance problem is explicitly acknowledged and discussed in both \citet{CS05} and \citet{primiceri05}. See also the discussion in \citet{CCM19}.} This ordering issue has become an increasingly pertinent problem due to two prominent developments in the VAR literature. First, in the last two decades there has been a gradual departure from conventional recursive or zero identification restrictions to other more credible identification schemes---such as identification by sign restrictions \citep{Faust98,CD02, Uhlig05}---that do not restrict the order of the variables. Despite this development, models of \citet{CS05} and \citet{primiceri05} continue to be used to first obtain reduced-form estimates, which are then taken as inputs in the subsequent structural analysis. Since the reduced-form estimates are not order invariant, the results from the structural analysis depend on the order of the variables in a subtle way, often without explicit recognition by the user.\footnote{The implications of this non-invariance problem for structural analysis have been illustrated in \citet{Bognanni18} and \citet{Hartwig2019}.}

Second, following the seminal contributions by \citet{BGR10} and \citet{koop13}, there is an increasing desire to use large VARs involving more than dozens of dependent variables for structural analysis. This development is partly motivated by the concern of informational deficiency of using a limited information set---by expanding the set of relevant variables, one can alleviate this concern \citep[see, e.g.,][]{HS91, LR93, LR94}. However, unless there is a natural variable ordering 
(e.g., using recursive identification restrictions), the ordering issue becomes more severe as the number of ways to order the variables increases exponentially with the number of variables. 

In view of these developments, we consider an alternative Bayesian VAR based on the factor stochastic volatility that is constructed to be invariant to the order of the dependent variables. Factor stochastic volatility models are commonly used for modeling high-dimensional financial data, but are less widely employed in empirical macroeconomics.\footnote{A notable exception is \citet{KH20}, who use Bayesian VARs with factor stochastic volatility for macroeconomic forecasting. \citet{CCM18} consider a related multiplicative 2-factor stochastic volatility model to study the impact of macroeconomic and financial uncertainty.} In specifying a suitable factor stochastic volatility model, there is often a tension between identification and order invariance. On the one hand, one can identify the factors and the associated factor loadings by fixing the orientation of the factors \citep[e.g., as in][]{GZ96,CNS06}. But this identification strategy essentially fixes the order of the variables, and therefore the identified model is not order invariant \citep[see, e.g., the discussion in][]{CLS18}. On the other hand, one could avoid fixing the orientation of the factors and obtain an order-invariant model, but it is unclear that the factors and the loadings are identified.\footnote{For example, \citet{Kastner19} does not impose any orientation restrictions on the factors, arguing that identification of the factor loadings is not necessary for his purpose of estimating the reduced-form covariance matrix.} 

We solve this dilemma between achieving identification and order invariance by carefully teasing out a set of conditions strong enough for identification, yet they are weak enough that the model remains order invariant. More specifically, we construct a VAR in which the innovations have a factor structure, and both the factors and the idiosyncratic errors follow stochastic volatility processes. We first show that the likelihood implied by this model is invariant to the order of the dependent variables. We then discuss sufficient conditions for identification of the factors and the factor loadings, building upon the approach in \citet{SF01} and extending it to a more general setting in which both the factors and the idiosyncratic errors are heteroscedastic. Under mild regularity conditions, we show that the factor loadings under our setup are identified up to permutation and sign changes. Furthermore, with additional sign restrictions that satisfy a set of conditions, we show that the factor loadings and the associated factors are point-identified. 

To determine the number of factors, we develop an estimator of the marginal likelihood based on an importance sampling approach to evaluate the observed-data or integrated likelihood. Through a series of Monte Carlo experiments, we show that our marginal likelihood estimator works well and is able to select the correct number of factors under a variety of settings.

We then discuss how our VAR with factor stochastic volatility (VAR-FSV) can be used for structural analysis. More specifically, we develop various structural analysis tools for VAR-FSV similar to those designed for standard structural VARs. In particular, we describe methods to construct structural impulse response functions, forecast error variance decompositions and historical decompositions. We demonstrate the methodology by revisiting the 6-variable VAR identified by a set of sign restrictions on the contemporaneous impact matrix considered in \citet{FRS19}. We augment their system to a 20-variable VAR by including additional, seemingly relevant macroeconomic and financial variables, which helps alleviate the concern of informational deficiency. In addition, the impulse responses obtained using the VAR-FSV with the sign restrictions imposed are point-identified. Empirically, we show that by including the additional variables and sign restrictions, one can substantially sharpen inference.

Our paper is related to the recent work by \citet{Korobilis20}, who uses a VAR with a factor error structure for structural analysis. His work is motivated by the computational challenge of imposing a large number of sign restrictions to obtain admissible draws using conventional accept-reject methods \cite[such as the widely used algorithm in][]{RWZ10}. This computational hurdle has so far limited the use of sign restrictions to relatively small systems with at most half a dozen dependent variables.\footnote{Large VARs, on the other hand, are mostly identified using recursive or zero restrictions. See, for example, \citet{LSZ96}, \citet{BGR10} and \citet{ER17}.}  Instead of using standard structural VARs, \citet{Korobilis20} assumes that the factors in his model play the role of structural shocks, and shows that in this case structural analysis can be done efficiently even when one imposes a large number of sign restrictions. His model, however, is homoscedastic, and consequently it is only set-identified. By contrast, in our VAR-FSV both the factors and the idiosyncratic errors follow stochastic volatility processes. This feature does not only accommodate the empirical finding that macroeconomic and financial variables typically exhibit time-varying volatility \citep[see, e.g.,][]{clark11, CR15}, it also allows us to achieve point-identification of the factors and the factor loadings.

Our work also contributes to the recent literature on using heteroscedasticity to identify conventional structural VARs, including \citet{WD15}, \citet{LLM10}, \citet{HL14}, \citet{BB20}, \citet{Lewis21} and \citet{BPSS21}. Our paper considers the alternative setting of a VAR with a factor stochastic volatility specification and establishes sufficient conditions for identification. One key advantage of using VAR-FSV for structural analysis, compared to structural VARs, is that under VAR-FSV it is computationally feasible to estimate large systems and impose a large number of sign restrictions. 

Our work is also related to the growing literature on constructing multivariate stochastic volatility models that are order invariant. One approach is based on Wishart or inverse-Wishart processes; examples include \citet{PG2006}, \citet{AM09}, \citet{CDLS18} and \citet{SZ20}. These models, however, are typically computationally intensive to estimate as the estimation involves drawing from non-standard high-dimensional distributions. As such, these models are generally not applicable to large datasets. An alternative approach is based on the common stochastic volatility models in \citet{CCM16} and \citet{chan20}. Although these models are designed for large systems and can be estimated quickly, they are more restrictive since the time-varying error covariance matrix depends on a single stochastic volatility process---in particular, the error variances are always proportional to each other. 

There are also order-invariant models that are based on the discounted Wishart process, such as those in \citet{Uhlig97}, \citet{WH06} and \citet{Bognanni18}. These models are convenient to estimate as they admit Kalman-filter type filtering and smoothing algorithms. The cost for this tractability, however, is that they are generally too tightly parameterized, and consequently, they tend to underperform in forecasting macroeconomic variables relative to standard stochastic volatility models such as \citet{CS05} and \citet{primiceri05} \citep[see][for an example]{ARRS21}. Lastly, the recent paper \citet{CKY21} extends the stochastic volatility model of \citet{CS05} by avoiding the use of Cholesky decomposition so that the extension is order-invariant. So far this reduced-form VAR is used for forecasting, and further research is needed to incorporate identification restrictions for structural analysis.

The rest of this paper is organized as follows. Section~\ref{s:VAR-FSV} first introduces the VAR with factor stochastic volatility. Its theoretical properties, including order invariance and sufficient conditions for identification, are discussed in Section \ref{s:properties}. We then outline a posterior sampler and a marginal likelihood estimator for the model in Section \ref{s:estimation} and Section~\ref{s:ML}, respectively. Next, Section~\ref{s:tools} develops various structural analysis tools for the VAR-FSV model, including algorithms to construct structural impulse response functions and to perform various decompositions. Then, Section~\ref{s:MC} presents Monte Carlo results to illustrate how well the marginal likelihood estimator works under a variety of settings. We next demonstrate the proposed methodology via a structural analysis with sign restrictions in Section~\ref{s:application}. Finally, Section~\ref{s:conclusion} concludes and discusses some future research directions.




\section{A Bayesian VAR with Factor Stochastic Volatility} \label{s:VAR-FSV}

In this section we outline a Bayesian VAR with factor stochastic volatility (FSV) and the associated prior distributions. To that end, let $\by_t$ be an $n\times 1 $ vector of dependent variables at time $t$. Then, for $t=1,\ldots, T$, consider the following VAR-FSV model:
\begin{align}
	\by_t       & = \ba_0 + \bA_1 \by_{t-1} + \cdots + \bA_p\by_{t-p} + \vepsilon_t,
	\label{eq:yt} \\
	\vepsilon_t & = \bL \mathbf{f}_t + \bu_t^y, 
	\label{eq:epsilont}
\end{align}
where $\mathbf{f}_t = (f_{1,t},\ldots, f_{r,t})'$ denotes a $r\times 1$ vector of latent factors and $\bL$ is an $n\times r$ matrix of factor loadings. 
Note also that $\bL$ is unrestricted. The disturbances $\bu_t^y$ and the latent factors $\mathbf{f}_t$ are assumed to be independent at all leads and lags. Moreover, they are specified as jointly Gaussian:
\begin{equation} \label{eq:ft}
	\begin{pmatrix}\bu_t^y \\  \mathbf{f}_t \end{pmatrix} \sim\distn{N}
	\left(\begin{pmatrix} \mathbf{0}\\ \mathbf{0} \end{pmatrix},
	\begin{pmatrix} \vSigma_t & \mathbf{0} \\ \mathbf{0} & \vOmega_t \end{pmatrix}\right),
\end{equation}
where $\vSigma_t = \text{diag}(\e^{h_{1,t}},\ldots, \e^{h_{n,t}})$ and $ \vOmega_t = \text{diag}(\e^{h_{n+1,t}},\ldots, \e^{h_{n+r,t}})$ are diagonal matrices. For $t=2,\ldots, T$, the log-volatilities evolve as:
\begin{align}
	h_{i,t} & = \mu_{i} + \phi_i(h_{i,t-1} - \mu_i) + u_{i,t}^h, \quad  
	u_{i,t}^h\sim\distn{N}(0,\sigma_i^2), \quad i=1,\ldots, n, \label{eq:ht1} \\
	h_{n+j,t} & = \phi_{n+j} h_{n+j,t-1} + u_{n+j,t}^h, \quad  
	u_{n+j,t}^h\sim\distn{N}(0,\sigma_{n+j}^2), \quad j=1,\ldots, r, \label{eq:ht2}
\end{align}
where we impose $|\phi_1|<1, \ldots |\phi_{n+r}|<1$ to ensure stationarity. Finally, the initial conditions follow the stationary distributions $h_{i,1} \sim \distn{N}(\mu_i,\sigma_i^2/(1-\phi_i^2)), i=1,\ldots, n,$ and $h_{n+j,1} \sim \distn{N}(0,\sigma_{n+j}^2/(1-\phi_{n+j}^2)), j=1,\ldots, r$. Note that the stationary distributions of the log-volatilities associated with the idiosyncratic errors have nonzero means, whereas the means of those associated with the factors are set to be zero for normalization.

To facilitate estimation, we rewrite the VAR in \eqref{eq:yt} as
\begin{equation} \label{eq:yt2}
	\by_t  =  (\mathbf{I}_n\otimes \bx_t')\vbeta + \vepsilon_t, 
\end{equation}
where $\mathbf{I}_n$ is the identity matrix of dimension $n$, $\otimes$ is the Kronecker product, $\vbeta = \text{vec}([\ba_0,  \bA_1, \ldots, \bA_p]')$ and $\bx_t = (1,\by_{t-1}',\ldots,\by_{t-p}')'$ is a $k \times 1$ vector of intercept and lagged values with $k=np+1$. 

Next, we specify the prior distributions on the model parameters. Let 
$\vbeta_i$ and $\bl_i$ denote the VAR coefficients and the elements of $\bL$ in the $i$-th equation, respectively, for $i=1,\ldots, n$. We assume the following independent priors on $\vbeta_i$ and $\bl_i$ for $i=1,\ldots, n$:
\[
	\vbeta_i\sim\distn{N}(\vbeta_{0,i},\bV_{\vbeta_i}),\quad \bl_i\sim\distn{N}(\bl_{0,i},\bV_{\bl_i}).
\]
We elicit the prior mean vector $\vbeta_{0,i}$ and the prior covariance matrix $\bV_{\vbeta_i}$ similar to the Minnesota prior \citep{DLS84, litterman86, KK93}. Specifically, for growth rates data, we set $\vbeta_{0,i} = \mathbf{0}$ to shrink the VAR coefficients to zero. For level data, $\vbeta_{0,i}$ is set to be zero as well except for the coefficient associated with the first own lag, which is set to be one. The prior covariance matrix 
$\bV_{\vbeta_i}$ is constructed so that it depends on two key hyperparameters, $\kappa_1$ and $\kappa_2$, that control respectively the overall shrinkage strength of `own' lags and `other' lags. For a more detailed discussion of the Minnesota prior, see, e.g., \citet{KK10}, \citet{DNS11} or \citet{karlsson13}. 

Finally, for the parameters in the stochastic volatility equations, we assume the priors:
\[
	\mu_i \sim \distn{N}(\mu_{0,i},V_{\mu_i}), \; \phi_j\sim \distn{N}(\phi_{0,j},V_{\phi_j})1(|\phi_j|<1),\; 	
	\sigma_{j}^2 \sim \distn{IG}(\nu_{j},S_{j}),
\]
$i=1,\ldots, n$ and $j=1,\ldots, n+r$.

\section{Order Invariance and Identification}  \label{s:properties}

In this section we describe a few important properties of the VAR-FSV model specified in \eqref{eq:yt}-\eqref{eq:ht2}. First, the likelihood implied by the model is invariant to the order of the variables (after permuting the relevant parameters appropriately). To see that, let $\bP$ be an $n\times n$ permutation matrix such that $\bP\bP' = \bP'\bP = \mathbf{I}_n$. For the $n$-variate Gaussian density $f_{\distn{N}}(\cdot ; \vmu,\vSigma)$ with mean vector $\vmu$ and covariance matrix $\vSigma$, it is easy to see that  $f_{\distn{N}}(\bx ; \vmu,\vSigma) = f_{\distn{N}}(\bP\bx ; \bP\vmu, \bP\vSigma\bP')$.

Next, we derive an expression of the likelihood function. To that end, stack $\bh_t^y = (h_{1,t},\ldots,h_{n,t})'$ and $\bh_t^f = (h_{n+1,t},\ldots,h_{n+r,t})'$. We similarly define $\vphi_y, \vphi_f, \vsigma^{2}_y $ and $\vsigma^{2}_f$. In addition, we let $\bh_t = (\bh_t^{y'},\bh_t^{f'})'$, $\vphi = (\vphi_y', \vphi_f')'$, $\vsigma^2 = (\vsigma^{2'}_y, \vsigma^{2'}_f)$ and $\vmu = (\mu_1,\ldots, \mu_n)'$. Then, the state equations \eqref{eq:ht1}-\eqref{eq:ht2} imply that the densities of $\bh_t^y$ and $\bh_t^f,$ for $t=2,\ldots, T$, are, respectively,  
\[
	f_{\distn{N}}(\bh_t^y; \vmu + \vphi_y\odot(\bh_{t-1}^y-\vmu),\text{diag}(\vsigma^{2}_y))
	\text{ and } f_{\distn{N}}(\bh_t^f; \vphi_f\odot \bh_{t-1}^f,\text{diag}( \vsigma^{2}_f)),
\]
where $\odot$ is the element-wise multiplication. Moreover, the initial conditions $\bh_1^y$ and $\bh_1^f$ have, respectively, the densities
\[
	f_{\distn{N}}(\bh_1^y; \vmu,\text{diag}(\vsigma^{2}_y\oslash(\mathbf{1} -  \vphi_y))), \text{ and }  
	f_{\distn{N}}(\bh_1^f; \mathbf{0},\text{diag}(\vsigma^{2}_f\oslash(\mathbf{1} -  \vphi_f))),
\]
where $\oslash$ denotes the element-wise division. 

Next, using the representation in \eqref{eq:yt2} and integrating out the factors, the density of $\by_t$ given the parameters and log-volatilities is $f_{\distn{N}}(\by_t; (\mathbf{I}_n \otimes \bx_t')\vbeta,\bL\vOmega_t\bL'+ \vSigma_t)$. Stacking $\by=(\by_1',\ldots, \by_T')'$, the likelihood function, or more precisely the integrated or observed-data likelihood, can therefore be written as
\begin{equation}\label{eq:like}
\begin{split}
	p(\by  \gvn\vbeta, & \bL,\vmu,\vphi,\vsigma^2)  =  \\
	 & \int f_{\distn{N}}(\bh_1^y; \vmu,\text{diag}(\vsigma^{2}_y\oslash(\mathbf{1} -  \vphi_y)))
	f_{\distn{N}}(\bh_1^f; \mathbf{0},\text{diag}(\vsigma^{2}_f\oslash(\mathbf{1} -  \vphi_f))) \\
	& \times \prod_{t=2}^T f_{\distn{N}}(\bh_t^y; \vmu + \vphi_y\odot(\bh_{t-1}^y-\vmu),\text{diag}(\vsigma^{2}_y))f_{\distn{N}}(\bh_t^f; \vphi_f\odot \bh_{t-1}^f,\text{diag}( \vsigma^{2}_f)) \\
	& \times \prod_{t=1}^T f_{\distn{N}}(\by_t; (\mathbf{I}_n \otimes \bx_t')\vbeta,\bL\vOmega_t\bL'+ \vSigma_t) \di \bh.
\end{split}
\end{equation}

Now, for an arbitrary permutation matrix $\bP$, suppose we permute the order of the dependent variables $\tilde{\by}_t = \bP\by_t$ and the associated lagged values $\tilde{\bx}_t' = (1,(\bP\by_{t-1})',\ldots, (\bP\by_{t-p})') = \bx_t'\bQ'$, where
$\bQ = \text{diag}(1,\mathbf{I}_p\otimes\bP)$. We claim that the likelihood implied by the VAR-FSV model is invariant to the permutation $\bP$ in the sense that 
\[
	p(\by \gvn\vbeta,\bL,\vmu,\vphi,\vsigma^2) = p(\tilde{\by} \gvn \tilde{\vbeta},\tilde{\bL},\tilde{\vmu},\tilde{\vphi},\tilde{\vsigma}^2),
\]
where $\tilde{\bL} = \bP\bL, \tilde{\vbeta} = (\bP\otimes\bQ) \vbeta, \tilde{\vmu} = \bP\vmu, \tilde{\vphi} = ((\bP\vphi_y)', \vphi_f')'$ and $\tilde{\vsigma}^{2} = ((\bP\vsigma_y^2)', \vsigma_f^{2'})'$.\footnote{Note that the permuted vector $\tilde{\vbeta}$ consists of the VAR coefficients of the following system stacked by rows:
\[
	\tilde{\by}_t  = \tilde{\ba}_0 +  \tilde{\bA}_1  \tilde{\by}_{t-1} + \cdots +  \tilde{\bA}_p 
	\tilde{\by}_{t-p} +  \tilde{\vepsilon}_t,
\]
where $ \tilde{\ba}_0 = \bP\ba_0$ and  $\tilde{\bA}_j = \bP\bA_j\bP', j=1,\ldots,p.$ That is,
\[
	\tilde{\vbeta} 
	= \text{vec}\begin{pmatrix} (\bP\ba_0)' \\ (\bP\bA_1\bP')' \\ \vdots \\ (\bP\bA_p\bP')'\end{pmatrix} 
	= \text{vec}\left( \bQ \begin{pmatrix} \ba_0' \\ \bA_1' \\ \vdots \\ \bA_p'\end{pmatrix} \bP'\right)
	= (\bP\otimes\bQ)\text{vec}\begin{pmatrix} \ba_0' \\ \bA_1' \\ \vdots \\ \bA_p'\end{pmatrix}
	= (\bP\otimes\bQ)\vbeta.
\]} That is, we obtain the same likelihood value for any permutation of $\by_t$ if the lagged values and the parameters are permuted accordingly. 
	
This claim of order invariance can be readily verified as follows. First, noting that 
\begin{align*}
	\bP(\mathbf{I}_n \otimes \bx_t')\vbeta & = (\bP\otimes 1)(\mathbf{I}_n \otimes \bx_t')(\bP'\otimes\bQ') (\bP\otimes\bQ) \vbeta \\	
	& = (\bP \mathbf{I}_n \bP') \otimes (1 \bx_t' \bQ')  (\bP\otimes\bQ) \vbeta \\
	& = (\mathbf{I}_n \otimes \tilde{\bx}_t')\tilde{\vbeta},
\end{align*}
we therefore obtain
\[
	f_{\distn{N}}(\by_t; (\mathbf{I}_n \otimes \bx_t')\vbeta,\bL\vOmega_t\bL'+ \vSigma_t) =
	f_{\distn{N}}(\tilde{\by}_t; (\mathbf{I}_n \otimes \tilde{\bx}_t')\tilde{\vbeta}, \tilde{\bL}\vOmega_t\tilde{\bL}' + \tilde{\vSigma}_t),
\]
where $\tilde{\vSigma}_t = \bP\vSigma_t\bP'$. Similarly, we also have
\begin{align*}
	 f_{\distn{N}}(\bh_1^y; \vmu,\text{diag}(\vsigma^{2}_y\oslash(\mathbf{1} -  \vphi_y)))
	& = f_{\distn{N}}(\bP\bh_1^y; \bP\vmu,\text{diag}((\bP\vsigma^{2}_y)\oslash(\mathbf{1} - \bP\vphi_y))),\\
	f_{\distn{N}}(\bh_t^y; \vmu + \vphi_y\odot(\bh_{t-1}^y-\vmu),\text{diag}(\vsigma^{2}_y))
	 & = f_{\distn{N}}(\bP\bh_t^y; \bP\vmu + (\bP\vphi_y)\odot(\bP\bh_{t-1}^y-\bP\vmu),
	\text{diag}(\bP\vsigma^{2}_y)).
\end{align*}

Since the Gaussian densities in \eqref{eq:like} are equal to their permuted counterparts, the integrand in $p(\tilde{\by} \gvn \tilde{\vbeta},\tilde{\bL},\tilde{\vmu},\tilde{\vphi},\tilde{\vsigma}^2)$ is exactly the same as that in \eqref{eq:like}. The only difference between the two integrals is the order of integration: $(\bh_t^y,\bh_t^f)$ versus $(\bP\bh_t^y,\bh_t^f)$. But since the integral is finite, one can change the order of integration without changing the integral. Hence, the desired result follows. The following proposition summarizes this result.
\begin{proposition}[Order Invariance] \label{thm:invar} \rm Let $p(\by \gvn\vbeta,\bL,\vmu,\vphi,\vsigma^2)$ denote the likelihood of the VAR-FSV model with lagged values $\bx_1,\ldots, \bx_T$. Let $\bP$ be an arbitrary $n\times n$ permutation matrix and define $\tilde{\by}_t = \bP\by_t$ and $\tilde{\bx}_t' = \bx_t'\bQ'$, where $\bQ = \text{diag}(1,\bP)$. Then, the VAR-FSV with dependent variables $\tilde{\by}_t $ and lagged values $\tilde{\bx}_t$ has the same likelihood. More precisely, 
\[
	p(\by \gvn\vbeta,\bL,\vmu,\vphi,\vsigma^2) = p(\tilde{\by} \gvn \tilde{\vbeta},\tilde{\bL},\tilde{\vmu},\tilde{\vphi},\tilde{\vsigma}^2),
\]
where $\tilde{\bL} = \bP\bL, \tilde{\vbeta} = (\bP\otimes\bQ) \vbeta, 
\tilde{\vmu} = \bP\vmu, \tilde{\vphi} = ((\bP\vphi_y)', \vphi_f')'$ 
and $\tilde{\vsigma}^{2} = ((\bP\vsigma_y^2)', \vsigma_f^{2'})'$. 
\end{proposition}

Next, we discuss sufficient conditions for identification of the factor loadings and latent factors. We mainly follow the approach in \citet{SF01}, but consider a more general setting in which the idiosyncratic errors $\bu_t^y$ in \eqref{eq:epsilont} are also heteroscedastic. First, note that it follows from \eqref{eq:epsilont} and \eqref{eq:ft} that the covariance matrix of $\vepsilon_t$ is given by $\var[\vepsilon_t \gvn \vOmega_t,\vSigma_t] =  \bL\vOmega_t\bL'+ \vSigma_t :=\vGamma_t$. The covariance structure of any observationally equivalent model to \eqref{eq:yt}--\eqref{eq:ft} with the same number of factors must satisfy $\vGamma_t =  \bL^* \vOmega_t^* \bL^{*'} + \vSigma_t^*$ for all $t$, where $ \bL^*$ is $n\times r$ and $\vOmega_t^*$ is $r\times r$. Furthermore, for a square matrix $\bA$ of dimension $m$, we define $\text{vecd}(\bA)$ to be the $m\times 1$ vector that stores its diagonal elements. Now, we consider the following assumptions that are used throughout the paper.

\begin{assumption} \label{ass1} \rm
The stochastic processes in $\text{vecd}(\vOmega_t)$ are linearly independent, i.e., there does not exist $\vdelta \in \mathbb{R}^{r}$, $\vdelta \neq \mathbf{0}$ such that  $\vdelta'\text{vecd}(\vOmega_t) = 0$ for all $t$. 
\end{assumption}

\begin{assumption} \label{ass2} \rm If any row of the matrix of factor loadings $\bL$ is deleted, there remain two disjoint submatrices of rank~$r$. 
\end{assumption}

Assumption~\ref{ass1} requires that no stochastic volatilities in the common factors can be expressed as a linear combination of other factor stochastic volatilities. Under our factor stochastic volatility model, this assumption is automatically satisfied. Assumption~\ref{ass2} limits the extent of sparseness in the factor loadings matrix to ensure one can separately identify the common and the idiosyncratic components. This assumption can be traced back to \citet{AR56}, and is widely adopted in the literature. Implicitly, it also requires that  
$r \leq (n-1)/2$. Since factor models are mostly applied to situations where the number of variables $n$ is much larger than the number of factors $r$, this is not a stringent condition. 




With Assumptions 1 and 2, one can show that the factor stochastic volatility model specified in \eqref{eq:epsilont}-\eqref{eq:ft} is identified up to permutations and sign changes of the factors. The identification results are summarized in the following proposition. 

\begin{proposition}[Identification of the Common Variance Component] \label{thm:iden1} \rm Under Assumption~1 and Assumption~2, the only observationally equivalent model to \eqref{eq:epsilont}-\eqref{eq:ht2} is ${\bL}^*={\bL}\bP_{r}\bP_{\pm}$, 
$\vOmega_t^* = \bP_{\pm}\bP_{r}\vOmega_t\bP_{r}'\bP_{\pm}'$ and ${\vSigma}_t^*={\vSigma}_t$, where $\bP_{r}$ is a permutation matrix of dimension $r$ and $\bP_{\pm}$ is a reflection matrix in which each diagonal entry is either~$+1$ or~$-1$.
\end{proposition}

We prove the proposition by adapting the results in \citet{AR56} and \citet{SF01} to our setting. The details are provided in Appendix~A. Proposition~\ref{thm:iden1} contains two sets of identification results. First, it shows that the common and idiosyncratic variance components can be separately identified. Second, the common variance component is identified up to permutations and sign switches of the latent factors. 

So far we have only considered the case where all $r$ factors are heteroskedastic. It turns out this is not necessary for identification of the common variance component. More generally, one can show that part of the factor loadings matrix is identified even when some of the factors are homoskedastic (their variances are normalized to be one). The following proposition summarizes such a partial identification result.  

\begin{proposition}[Partial Identification of the Common Variance Component When the Number of Heteroskedastic Factors Is $r_1 < r$] \label{thm:iden2} \rm Let $\vOmega_t = \diag(\vOmega_{1t},\mathbf{I}_{r_2})$, where $\vOmega_{1t}$ is a $r_1\times r_1$ covariance matrix and $\mathbf{I}_{r_2}$ is the $r_2$-dimensional identity matrix with $r=r_1+r_2$. Similarly partition $\bL = (\bL_1, \bL_2)$ such that $\bL_1$ is $n\times r_1$ and $\bL_2$ is $n\times r_2$.  If $\text{diag}(\vOmega_{1t},1)$ satisfies Assumption 1 and $\bL$ satisfies Assumption~2, then $\bL_1$ is identified up to permutations and sign switches.
\end{proposition}

The proof of this proposition is given in Appendix~A. The condition that $\text{diag}(\vOmega_{1t},1)$ satisfies Assumption 1---i.e., $(\text{vecd}(\vOmega_{1t})', 1)'$ are linearly independent for all $t$---requires all stochastic processes in $\text{vecd}(\vOmega_{1t})$ to be non-degenerate. (Otherwise those homoskedastic factors should be relocated  to the homoskedastic part.) It is also worth noting that Proposition~\ref{thm:iden2} does not imply that for $r_1<r$, the common variance component is not identifiable. In fact, it turns out that the minimum number of heteroskedastic factors for identifying $\bL$ (up to permutations and sign switches) is $ r_1 =  r-1$. We summarize this result in the following corollary.

\begin{corollary} \label{thm:coro1} \rm Under the assumptions in Proposition~\ref{thm:iden2}, if the number of heteroskedastic factors is $r_1\geq r-1$, then  $\bL$ is identified up to permutations and sign switches.
\end{corollary}

The reason why only $r_1 = r-1$ heteroskedastic factors are needed for identification is intuitive. Under the assumptions in Proposition~\ref{thm:iden2}, when $r_1 = r-1$, only one element in $\vOmega_t$ is normalized to one; the remaining $r-1$ stochastic processes in $\vOmega_{1t}$ are linearly independent. Consequently, $\vOmega_t$ also satisfies Assumption 1. And Corollary~\ref{thm:coro1} follows from Proposition~\ref{thm:iden1}.


For $r_1 < r-1$, part of the factor loadings matrix $\bL$ is invariant under general orthogonal transformation. To see that, suppose $r_1 < r-1$, and hence $\bL_2$ has at least $r_2 \geq 2$ columns. Let $\bR_{f_2}$ be a $r_2\times r_2$ orthogonal matrix other than permutation $\bP_{r_2}$ or reflection $\bP_{\pm}$ such that $\bR_{f_2}\bR_{f_2}' = \mathbf{I}_{r_2}$.\footnote{The only one-dimensional orthogonal matrices are reflections, namely, $+1$ and $-1$. Hence, $r_2$ must be at least 2.} Then, we have $\bL\vOmega_t\bL' = \bL_1\vOmega_{1t}\bL_1' + \bL_2\bL_2' = \bL_1\vOmega_{1t}\bL_1' + \bL_2\bR_{f_2} \bR_{f_2}'\bL_2'.$ Hence, $\bL^* = (\bL_1,\bL_2\bR_{f_2})$ and $\vOmega^*_t = \vOmega_t$ form an observationally equivalent model. 

For point-identification, one needs additional restrictions on $\bL$ (or the latent factors) to pin down the unique permutation and sign configuration. As is common in macroeconomic analysis using VARs, sign restrictions implied by economic theory are often available to assist structural identification. For a recent contribution linking sign restrictions and factor models, see \citet{Korobilis20}. Below we describe how we can incorporate sign restrictions to achieve point-identification. To that end, let $\bS$ denote the $n\times r$ matrix that collects the corresponding restrictions on the factor loadings matrix $\bL$. The entries of $\bS$ can take four values: 1, $-1$, 0 and N/A, which denotes positive restriction, negative restriction, zero restriction and no restrictions, respectively. For example, if economic theory implies that $\bL_{ij}>0 $, then $\bS_{ij} = +1$; if there are no restrictions on $\bL_{ij} $, then $\bS_{ij} = \text{N/A}$.

Recall that under Assumptions 1-2, Proposition~\ref{thm:iden1} dictates that the factor loadings matrix $\bL^*$ of any observationally equivalent model must be of the form $\bL^* = \bL\bP$, where $\bP$ is a product of a reflection and a permutation. To be observationally equivalent with the sign restrictions imposed---i.e., satisfying exactly the same sign restrictions---we must have $\bS\bP =  \bS$. Intuitively then, for point-identification of $\bL$ there must be enough structure in $\bS$ such that the only possible $\bP$ is the identify matrix. Now, suppose that each column of $\bS$ has at least one sign restriction and no columns are the same or negative of any other columns. These conditions are sufficient as they rule out any permutations or sign changes except the identity.

To see that the conditions are necessary, suppose there is a column that has no sign restrictions. Then, changing the sign of the associated column in $\bL$ (and the associated rows in $\bff_t$) would leave the model observationally equivalent. Next, suppose one column is the same or the negative of any other column, then we can permute (and change signs if necessary) the relevant columns to leave the model observationally equivalent. We summarize these results in the following corollary.

\begin{corollary} \label{thm:coro2} \rm Under Assumptions 1-2, the necessary and sufficient conditions for point-identification of the factor loadings matrix are that each column of $\bS$ has at least one sign restriction and no columns are the same or negative of any other columns.
\end{corollary}

\section{Bayesian Estimation} \label{s:estimation}

In this section we describe an efficient posterior sampler to estimate the model in \eqref{eq:yt}--\eqref{eq:ht2} with signs or zero restrictions specified in $\bS$. Below we note a few details in our implementation with the goal of improving speed and sampling efficiency. First, even though the factors $\bff_1,\ldots, \bff_T$ are conditionally independent given the data and other parameters, we sample them jointly in one step using the precision sampler of \citet{CJ09}---instead of drawing them sequentially in a for-loop---to speed up the computations. 

Second, since VARs tend to have a lot of parameters even for small and medium systems, we implement an equation-by-equation estimation approach similar in spirit to that in \citet{CCM19} to sample the VAR coefficients. Specifically, given the latent factors $\mathbf{f}$, the VAR becomes $n$ unrelated regressions, and one can sample the VAR coefficients equation by equation without any loss of efficiency. Third, with the sign restrictions imposed in $\bS$, the full conditional distribution of the factor loadings in each equation becomes a truncated multivariate normal distribution. To sample from such a distribution, we use the algorithm in \citet{botev17} that is based on quadratic programming.

For notational convenience, stack $\by = (\by_1',\ldots, \by_T')',$ $\mathbf{f} = (\mathbf{f}_1',\ldots, \mathbf{f}_T')',$ $\bh = (\bh_1',\ldots, \bh_{T}')'$ and $\vbeta = (\vbeta_1',\ldots, \vbeta_n')'$. In addition, let $\by_{i,\bigcdot} = (y_{i,1},\ldots, y_{i,T})'$ denote the vector of observed values for the 
$i$-th variable, $i=1,\ldots, n$. We similarly define $\bh_{i,\bigcdot} = (h_{i,1},\ldots, h_{i,T})', 
i=1,\ldots, n+r$. Then, posterior draws can be obtained by sampling sequentially from: 
\begin{enumerate}
	\item $p(\mathbf{f} \gvn \by, \vbeta, \bL, \bh, \vmu, \vphi,\vsigma^2) 
	= p(\mathbf{f} \gvn \by, \vbeta, \bL, \bh)$; 
	
	\item $p(\vbeta,\bL \gvn \by, \mathbf{f}, \bh, \vmu, \vphi,\vsigma^2) = 
	\prod_{i=1}^n p(\vbeta_i,\bl_i \gvn \by_{i,\bigcdot}, \mathbf{f}, \bh_{i,\bigcdot})$; 
	
	\item $p(\bh \gvn \by, \mathbf{f},\vbeta, \bL, \vmu, \vphi,\vsigma^2) = 
	\prod_{i=1}^{n+r} p(\bh_{i,\bigcdot} \gvn \by, \mathbf{f}, \vbeta, \bL, \vmu, \vphi,\vsigma^2)$; 
	
	\item $p(\vsigma^2 \gvn \by, \mathbf{f}, \vbeta, \bL, \bh, \vmu, \vphi) 	
	= \prod_{i=1}^{n+r} p(\sigma_i^2 \gvn \bh_{i,\bigcdot}, \mu_i, \phi_i)$;
	
	\item $p(\vmu \gvn \by, \mathbf{f}, \vbeta, \bL, \bh, \vphi,\vsigma^2) 
	= \prod_{i=1}^{n} p(\mu_i \gvn \bh_{i,\bigcdot}, \phi_i, \sigma^2_i) $; 
	
	\item $p(\vphi \gvn \by, \mathbf{f}, \vbeta, \bL, \bh, \vmu,\vsigma^2) 
	= \prod_{i=1}^{n+r} p(\phi_i \gvn \bh_{i,\bigcdot}, \mu_i, \sigma^2_i) $.	
	
\end{enumerate}

\textbf{Step 1}. As mentioned above, since the factors $\bff_1,\ldots, \bff_T$ are conditionally independent given other parameters, in principle one can sample each factor sequentially in a for-loop. Here, however, we vectorize all the operations and sample them jointly in one step to improve computational speed. More specifically, we first stack the VAR in \eqref{eq:yt}-\eqref{eq:epsilont} over $t=1,\ldots, T$ and write it as:
\[
	\by = \bX\vbeta + (\mathbf{I}_T\otimes \bL)\mathbf{f} +  \bu^y, \quad \bu^y \sim \distn{N}(\mathbf{0},\vSigma),
\]	
where $\bX$ is the matrix of intercepts and lagged values and $\vSigma = \text{diag}(\vSigma_1,\ldots, \vSigma_T)$
with $\vSigma_t = \text{diag}(\e^{\bh_t^y})$. In addition, it follows from \eqref{eq:ft} that $(\mathbf{f} \gvn \bh)\sim \distn{N}(\mathbf{0},\vOmega),$ where $\vOmega = \text{diag}(\vOmega_1,\ldots, \vOmega_T)$ with $\vOmega_t = \text{diag}(\e^{\bh_t^f})$.

Then, by standard linear regression results \citep[see, e.g.,][chapter 12]{CKPT19}, we have
\begin{equation}\label{eq:postf}
	(\mathbf{f} \gvn\by, \vbeta, \bL, \bh) \sim  \distn{N}(\hat{\mathbf{f}},\bK_{\mathbf{f}}^{-1}), 
\end{equation}
where
\begin{equation} \label{eq:Kf}
	\bK_{\mathbf{f}} = \vOmega^{-1} + (\mathbf{I}_T\otimes \bL')\vSigma^{-1}(\mathbf{I}_T\otimes \bL), \quad 
	\hat{\mathbf{f}}  = \bK_{\mathbf{f}}^{-1}(\mathbf{I}_T\otimes \bL')\vSigma^{-1}(\by-\bX\vbeta).
\end{equation}
Note that the precision matrix $\bK_{\mathbf{f}}$ is a band matrix, i.e., it is sparse and all the nonzero entries are arranged along the diagonal bands above and below the main diagonal. As such, once can use the precision sampler of \citet{CJ09} to sample $\mathbf{f}$ efficiently.

\textbf{Step 2}. Next, we sample $\vbeta$ and $\bL$ jointly to improve sampling efficiency. Given the latent factors $\mathbf{f}$, the VAR becomes $n$ unrelated regressions and we can sample $\vbeta$ and $\bL$ equation by equation. Recall that $\by_{i,\bigcdot} = (y_{i,1},\ldots, y_{i,T})'$ is defined to be the $T\times 1$ vector of observations for the $i$-th variable; and that $\vbeta_i$ and $\bl_i$ represent, respectively, the VAR coefficients and the factor loadings in the $i$-th equation. Then, the $i$-th equation of the VAR can be written as
\[
	\by_{i,\bigcdot} = \bX_i\vbeta_i + \mathbf{F} \bl_i + \bu_{i,\bigcdot}^y,
\]
where $\mathbf{F} = (\mathbf{f}_{1,\bigcdot},\ldots, \mathbf{f}_{r,\bigcdot})$ is the $T\times r$ matrix of factors with $\mathbf{f}_{i,\bigcdot} = (f_{i,1},\ldots, f_{i,T})'$. The vector of disturbances $\bu_{i,\bigcdot}^y= (u_{i,1},\ldots, u_{i,T})'$ is distributed as $\distn{N}(\mathbf{0},\vOmega_{\bh_{i,\bigcdot}})$, where $\vOmega_{\bh_{i,\bigcdot}} =\text{diag}(\e^{h_{i,1}},\ldots, \e^{h_{i,T}})$.\footnote{Note that zero restrictions on $ \bl_i$ can be easily handled by redefining $\bl_i $ and $ \mathbf{F}$ appropriately. For example, if the first element of $\bl_i $ is restricted to be zero, we can define 
$\tilde{\bl}_i $ to be the vector consisting of the second to $r$-th elements of $\bl_i$ and $\tilde{\mathbf{F}} = (\mathbf{f}_{2,\bigcdot},\ldots, \mathbf{f}_{r,\bigcdot})$. Then, we replace  $\mathbf{F} \bl_i$
 by  $\tilde{\mathbf{F}}\tilde{\bl}_i$.} Letting $\vtheta_i = (\vbeta_i',\bl_i')'$ and  $\bZ_i = (\bX_i,\mathbf{F})$, we can further write the VAR systems as
\[
	\by_{i,\bigcdot} =\bZ_i\vtheta_i + \bu_{i,\bigcdot}^y.
\]
Let $R_i\subset \mathbb{R}^r$ be the support of $\bl_i$ defined by the sign restrictions specified in the $i$-th row of $\bS$. Then, using standard linear regression results, we obtain: 
\[
	(\vtheta_i \gvn \by_{i,\bigcdot}, \mathbf{f}, \bh_{i,\bigcdot}) \sim  \distn{N}(\hat{\vtheta}_i,\bK_{\vtheta_i}^{-1})1(\bl_i\in R_i), 
\]
where
\[
	\bK_{\vtheta_i} = \bV_{\vtheta_i}^{-1} + \bZ_i'\vOmega_{\bh_{i,\bigcdot}}^{-1}\bZ_i, \quad 
	\hat{\vtheta}_i  = \bK_{\vtheta_i}^{-1}(\bV_{\vtheta_i}^{-1}\vtheta_{0,i} +  \bZ_i\vOmega_{\bh_{i,\bigcdot}}^{-1}
	\by_{i,\bigcdot}) 
\]
with $\bV_{\vtheta_i} = \text{diag}(\bV_{\vbeta_i},\bV_{\bl_i})$ and $\vtheta_{0,i} = (\vbeta_{0,i}',\bl_{0,i}')'$. A draw from this truncated multivariate normal distribution can be obtained using the algorithm in \citet{botev17}. The remaining steps are standard and we leave the details to Appendix B. Some simulation results are reported in Appendix D to show that the posterior sampler works well and the posterior estimates track the true values closely.

It is worth noting that Proposition~\ref{thm:iden1} only guarantees that the factors and factor loadings are identified up to permutations and sign changes. Hence, in practice one might encounter the so-called label-switching problem. One way to handle this issue is to post-process the posterior draws to sort them into the correct categories; see, e.g., \citet{KS19} for such an approach. In our empirical application we impose sign restrictions that satisfy Corollary \ref{thm:coro2}---and consequently the factors and factor loadings are point-identified. 

Next, we document the runtimes of estimating the VAR-FSV of different dimensions to assess how well the posterior sampler scales to higher dimensions. More specifically, we report in Table~\ref{tab:times} the computation times (in minutes) to obtain 10,000 posterior draws from the VAR-FSV of dimensions $n= 15, 30, 50$ and sample sizes $T=300, 800$. The posterior sampler is implemented in $\matlab$ on a typical desktop with an Intel Core i5-9600 @3.10 GHz processor and 16 GB memory. It is evident from the table that even for high-dimensional applications with 50 variables, the VAR-FSV model with sign restrictions imposed on the factor loadings can be estimated fairly quickly.

\begin{table}[H]
\centering
\caption{The computation times (in minutes) to obtain 10,000 posterior draws from the VAR-FSV model with $n$ variables and $T$ observations. All VARs have $r=4$ factors and $p = 4$ lags.} \label{tab:times}
\begin{tabular}{cccccc}
\hline \hline
 \multicolumn{3}{c}{$T = 300$} & \multicolumn{3}{c}{$T = 800$} \\
     $n = 15$  & $n=30$  & $n=50$  & $n = 15$  & $n=30$  & $n=50$  \\ \hline
     12.5  & 25.7 & 45.0 & 33.3 & 67.3 & 110.2 \\ \hline \hline
\end{tabular}
\end{table}

\section{Bayesian Model Comparison} \label{s:ML}

This section first gives a brief overview on the theory of Bayesian model comparison via the marginal likelihood. Then, we introduce an algorithm to evaluate the likelihood, or more precisely the integrated likelihood marginal of the latent states, implied by the VAR-FSV model. Finally, we present an adaptive importance sampling algorithm to estimate the marginal likelihood under the VAR-FSV model.

Suppose we wish to compare a collection of models $\{M_{1},\ldots, M_{K} \}$, where
each model $M_k$ is defined by a likelihood function $p(\by\gvn \vtheta_k, M_{k})$ and a prior on the model-specific parameter vector $\vtheta_k$ denoted by $p(\vtheta_k \gvn M_k)$. The gold standard for Bayesian model comparison is the Bayes factor in favor of $M_i$ against $M_j$, defined as
\[
	\text{BF}_{ij}  = \frac{p(\by\gvn M_i)}{p(\by\gvn M_j)},
\]
where
\begin{equation}\label{eq:ml}
	p(\by\gvn M_{k}) = \int p(\by\gvn \vtheta_k, M_{k}) p(\vtheta_k\gvn M_{k})\di\vtheta_k
\end{equation}
is the \emph{marginal likelihood} under model $M_k$, $k=i,j.$ This Bayes factor is related to the posterior odds ratio between the two models:
\[
	\frac{\Pm(M_i\gvn\by)}{\Pm(M_j\gvn\by)} = \frac{\Pm(M_i)}{\Pm(M_j)}\times \text{BF}_{ij},
\]
where $\Pm(M_i)/\Pm(M_j)$ is the prior odds ratio. It if clear that if both models are equally probable \textit{a priori}, i.e., $p(M_i) = p(M_j)$, then the posterior odds ratio between the two models is equal to the Bayes factor. Hence, the Bayes factor has a natural interpretation and is easy to understand. For example, under equal prior odds, if $\text{BF}_{ij} = 50$, then model $M_i$ is 50 times more likely than model $M_j$ given the data. For a more detailed discussion of the Bayes factor and its role in Bayesian model comparison, see \citet{koop03} or \citet{CKPT19}. From here onwards we suppress the model indicator.

\subsection{Integrated Likelihood Evaluation} \label{ss:intlike}

To estimate the marginal likelihood, we first present an efficient way to evaluate the likelihood, or more precisely the integrated likelihood marginal of the latent states, given in~\eqref{eq:like} . For notational convenience, we rewrite~\eqref{eq:like} as
\begin{equation}\label{eq:like2}
	p(\by \gvn\vbeta, \bL,\vmu,\vphi,\vsigma^2)  = \int  p(\by\gvn\vbeta,\bL,\bh) p(\bh \gvn \vmu,\vphi,\vsigma^2) 
	\di\bh,
\end{equation}
where the conditional density of $\by$ given $\bh$ but marginal of $\bff$ has the explicit expression
\[
	p(\by\gvn\vbeta,\bL,\bh) = (2\pi)^{-\frac{Tn}{2}}\prod_{t=1}^T|\bL\vOmega_t\bL'+\vSigma_t|^{- \frac{1}{2}}
	\e^{-\frac{1}{2}(\by_t - (\mathbf{I}_n\otimes\bx_t')\vbeta)'(\bL\vOmega_t\bL'+\vSigma_t)^{-1}
	(\by_t - (\mathbf{I}_n\otimes\bx_t')\vbeta)}.
\]
The second term of the integrand, $p(\bh \gvn \vmu,\vphi,\vsigma^2)$, is a $T(n+r)$-variate Gaussian density implied by the state equations specified in \eqref{eq:ht1}-\eqref{eq:ht2}. Its analytical expression is given in Appendix~C, and in particular, its precision matrix is banded. Hence, both densities can be evaluated quickly. 

The main difficulty in evaluating the integrated likelihood in \eqref{eq:like2}, however, is that it requires integrating out all the latent log-volatilities, which involves solving a $T(n+r)$-dimensional integral. In what follows, we adopt the importance sampling approach developed for time-varying parameter VARs in \citet{CE18} to our setting.\footnote{There is a long tradition of using importance sampling to evaluate the integrated likelihood of stochastic volatility models. Earlier papers, such as \citet{DK97}, \citet{SP97}, \citet{KH02}, \citet{FW08}, \citet{McCausland12}, have focused mostly on univariate stochastic volatility models.} More specifically, given an importance sampling density $g$---that might depend on model parameters and the data---we evaluate the integrated likelihood via importance sampling:
\begin{equation}\label{eq:IS_intlike}
	\hat{p}(\by \gvn \vbeta, \bL,\vmu,\vphi,\vsigma^2) = \frac{1}{R_1}
	\sum_{r=1}^{R_1}\frac{p(\by\gvn\vbeta,\bL,\bh^{(r)})p(\bh^{(r)}\gvn \vmu,\vphi,\vsigma^2)}
		{g(\bh^{(r)}; \by, \vbeta, \bL,\vmu,\vphi,\vsigma^2)},
\end{equation}
where $\bh^{(1)},\ldots, \bh^{(R_1)}$ are independent draws from $g$.

The choice of the importance sampling density $g$ is vital as it determines the variance of the estimator.
In general, we would like to use an importance sampling density so that it well approximates the integrand in \eqref{eq:like2}. Our particular choice is motivated by the observation that there is, in fact, a theoretical zero-variance importance sampling density---it is $p(\bh\gvn\by,\vbeta,\bL,\vmu,\vphi,\vsigma^2)$, the conditional posterior distribution of $\bh$ given the other parameters but marginal of $\bff$. In practice, however, this density cannot be used as an importance sampling density as it is non-standard (e.g., its normalizing constant is unknown and it is unclear how one can efficiently generate samples from this density). But this observation provides us guidance for selecting a good importance sampling density.

In particular, we aim to approximate this ideal importance sampling density using a Gaussian density. This is accomplished as follows. We first develop an expectation-maximization (EM) algorithm to locate the mode of $\log p(\bh\gvn\by,\vbeta,\bL,\vmu,\vphi,\vsigma^2)$, denoted as 
$\hat{\bh}$. Then, we obtain the negative Hessian of this log-density evaluated at the mode, denoted as $\bK_{\bh}$. The mode and the negative Hessian are then used, respectively, as the mean vector and precision matrix of the Gaussian approximation. That is, the importance sampling density is $\distn{N}(\hat{\bh},\bK_{\bh}^{-1}).$ We leave the technical details to Appendix~C. Below we comment on a few computational details. 

First, in the M-step of the EM algorithm, one needs to solve a $T(n+r)$-dimensional maximization problem, which is in general extremely computationally intensive. In our case, however, we are able to obtain analytical expressions of the gradient and the Hessian of the objective function (i.e., the Q-function), which allows us to implement the Newton-Raphson method. Furthermore, one can show that the Hessian is a) negative definite anywhere in $\mathbb{R}^{T(n+r)}$, and b) a band matrix. The former property guarantees rapid convergence of the Newton-Raphson method, while the latter property substantially speeds up the computations.

Second, to construct the importance sampling estimator in \eqref{eq:IS_intlike}, one needs to both evaluate and sample from the $T(n+r)$-dimensional Gaussian importance sampling density $M$ times. For very high-dimensional Gaussian densities, both operations are generally computational costly. For our Gaussian importance sampling density, however, we can show that its precision matrix is banded. As such, samples from this Gaussian density can be obtained quickly using the precision sampler in \citet{CJ09}. Evaluation of the density can be done just as quickly. We summarize the evaluation of the integrated likelihood in Algorithm \ref{alg:intlike}.

\begin{algorithm}[H]
\caption{Integrated likelihood estimation.}
\label{alg:intlike}
Given the parameters $\vbeta$, $\bL,$ $\vmu,\vphi$ and $\vsigma^2$, complete the following two steps.
\begin{enumerate}
	\item Obtain the mean vector $\hat{\bh}$ and precision matrix $\bK_{\bh}$ of the Gaussian importance sampling density detailed in Appendix C.

	\item For $r = 1,\ldots, R_1$, simulate $\bh^{(r)} \sim \distn{N}(\hat{\bh},\bK_{\bh}^{-1})$ using 
	the precision sampler in \citet{CJ09}, and compute the average
	\[
		\hat{p}(\by \gvn \vbeta, \bL,\vmu,\vphi,\vsigma^2) = \frac{1}{R_1}
	\sum_{r=1}^{R_1}\frac{p(\by\gvn\vbeta,\bL,\bh^{(r)})p(\bh^{(r)}\gvn \vmu,\vphi,\vsigma^2)}
		{g(\bh^{(r)}; \by, \vbeta, \bL,\vmu,\vphi,\vsigma^2)}.
	\]	
\end{enumerate}
\end{algorithm}

\subsection{Marginal Likelihood Estimation}

Next, we discuss the marginal likelihood estimation of the VAR-FSV model using an adaptive importance sampling approach called the improved cross-entropy method. This method requires little explicit analysis from the user and is applicable to a wide variety of problems (in contrast to the importance sampling estimator of the integrated likelihood estimation presented in Algorithm \ref{alg:intlike} that requires a lot of analysis). More specifically, suppose we wish to estimate the marginal likelihood $p(\by)\equiv p(\by\gvn M_k)$ given in~\eqref{eq:ml} using the following importance sampling estimator:
\begin{equation} \label{eq:ISml}
    \widehat{p(\by)}_{\rm IS} = \frac{1}{R_2}\sum_{r=1}^{R_2}
		\frac{p(\by\gvn \vect{\theta}^{(r)})p(\vect{\theta}^{(r)})}{g(\vect{\theta}^{(r)})},
\end{equation}
where $\vect{\theta}^{(1)},\ldots, \vect{\theta}^{(R_2)}$ are independent draws from the importance sampling density $g(\cdot)$. In particular, for our FSV model, $\vtheta = \{\vbeta,\bL,\vsigma^2,\vmu,\vphi\}$. While this importance sampling estimator in theory is unbiased and simulation consistent for any $g$---as long as it dominates $p(\by\gvn\cdot)p(\cdot)$, i.e., $g(\vtheta)=0\Rightarrow p(\by\gvn\vtheta)p(\vtheta)=0$---in practice its performance heavily depends on the choice of~$g$. Here we follow \citet{CE15} to use the improved cross-entropy method to construct $g$ optimally.\footnote{The original cross-entropy method was developed for rare-event simulation by \citet{rubinstein97, rubinstein99} using a multi-level procedure to construct the optimal importance sampling density. \citet{CE15} later show that this optimal importance sampling density can be obtained more accurately in one step using Markov chain Monte Carlo methods.}

To motivate the improved cross-entropy method, first note that the ideal zero-variance importance sampling density is the posterior density $p(\vtheta \gvn\by)$. That is, if we use $g^*(\vtheta) = p(\vtheta\gvn\by) = p(\by\gvn\vtheta)p(\vtheta)/p(\by)$ as the importance sampling density, then the associated estimator in~\eqref{eq:ISml} has zero variance. Unfortunately, $g^*$ cannot be used in practice as its normalization constant is precisely the marginal likelihood, the unknown quantity we aim to estimate. This nevertheless provides a benchmark to construct an optimal importance sampling density. More specifically, we aim to find a density that is `close' to this benchmark $g^*$ that can be used as an importance sampling density. 

To that end, consider a parametric family $\mathcal{G} = \{ g(\vect{\theta};\bv) \}$
indexed by the parameter vector $\bv$. We then find the density $g(\vtheta;\bv^*)\in\mathcal{G}$ such that it is, in a well-defined sense, the `closest' to $g^*$. One convenient measure of closeness between densities is the Kullback-Leibler divergence or the cross-entropy distance. More precisely, for two density functions $g_1$ and $g_2$, the cross-entropy distance from $g_1$ to $g_2$ is defined as:
\[
	\mathcal{D}(g_1,g_2) = \int g_1(\bx)\log \frac{g_1(\bx)}{g_2(\bx)}\di\bx.
\]
Given this measure of closeness, we obtain the density $g(\cdot;\bv)\in\mathcal{G}$
such that $\mathcal{D}(g^*,g(\cdot;\bv))$ is minimized, i.e., 
$\bv_{\text{ce}}^* = \argmin_{\bv}\mathcal{D}(g^*,g(\cdot;\bv))$. It can be shown that
solving the CE minimization problem is equivalent to finding
\[
	\bv^*_{\text{ce}} = \argmax_{\mathbf{v}}\int p(\by\gvn\vtheta)p(\vtheta)\log g(\vtheta;\bv)\di\vtheta.
\]

In general this optimization problem is difficult to solve analytically as it involves a high-dimensional integral. Instead, we consider its stochastic counterpart:
\begin{equation} \label{eq:maxMC}
	\widehat{\bv}^*_{\text{ce}} = \argmax_{\bv} \frac{1}{M}\sum_{m=1}^M \log g(\vect{\theta}_m; \bv),
\end{equation}
where $\vect{\theta}_1,\ldots, \vect{\theta}_M$ are posterior draws from 
$p(\vtheta\gvn\by) \propto p(\by\gvn\vtheta)p(\vtheta)$. It is useful to note that 
$\widehat{\bv}^*_{\text{ce}}$ is exactly the maximum likelihood estimate for $\bv$ if we view 
$g(\vect{\theta};\bv)$ as the likelihood function with parameter vector $\bv$ and $\vtheta_1,\ldots, \vtheta_M$ as an observed sample. Since finding the maximum likelihood estimator is a standard problem, solving~\eqref{eq:maxMC} is typically easy. For example, analytical solutions to \eqref{eq:maxMC} are available for the exponential family \citep[e.g.,][p. 70]{rk:ce}.

Next, we discuss the choice of the parametric family $\mathcal{G}$. One convenient class of densities is one in which each member $g(\vtheta ; \bv)$ is a product of probability densities, e.g., $g(\vtheta; \bv) = g(\vtheta_1; \bv_1)\times\cdots\times g(\vtheta_B; \bv_B)$, where $\vtheta = \{\vtheta_1,\ldots, \vtheta_B\}$ and $\bv = \{\bv_1,\ldots,\bv_B\}$. One main advantage of this choice is that we can then reduce the generally high-dimensional maximization problem~\eqref{eq:maxMC} into $B$ separate low-dimensional maximization problems. For example, for the our FSV model, we divide $\vtheta = \{\vbeta,\bL,\vsigma^2,\vmu,\vphi\}$ into 5 natural blocks, and consider the parametric family
\[
\begin{split}
	\mathcal{G} = & \left\{ g_{\distn{N}}(\vbeta; \bv_{1,\vbeta}, \bv_{2,\vbeta})
	 g_{\distn{N}}(\bL; \bv_{1,\bL}, \bv_{2,\bL})
	\prod_{i=1}^{n+r} g_{\distn{IG}}(\sigma_{i}^2; v_{1,\sigma_i^2}, v_{2,\sigma_i^2})
	\prod_{i=1}^{n}g_{\distn{N}}(\mu_{i}^2; v_{1,\mu_i}, v_{2,\mu_i})	\right. \\
	& \quad \times \left. 		
	\prod_{i=1}^{n+r}g_{\distn{N}}(\phi_i; v_{1,\phi_i}, v_{2,\phi_i})1\left(|\phi_i|<1\right)
	\right\},
\end{split}
\]
where $g_{\distn{N}}$ and $g_{\distn{IG}}$ are, respectively, the Gaussian and the inverse-gamma densities. Given this choice of the parametric family, the maximization problem in \eqref{eq:maxMC} can be readily solved (either analytically or using numerical optimization). 

Given the optimal importance density, denoted as $g(\vbeta,\bL,\vsigma^2,\vmu,\vphi; \bv^*)$, we construct the following importance sampling estimator:
\begin{equation}\label{eq:IS_ml}
	\widehat{p(\by)} = \frac{1}{R_2}
	\sum_{r=1}^{R_2}\frac{p(\by\gvn \vbeta^{(r)},\bL^{(r)},\vsigma^{2(r)},\vmu^{(r)},\vphi^{(r)})
	p(\vbeta^{(r)},\bL^{(r)},\vsigma^{2(r)},\vmu^{(r)},\vphi^{(r)})}
		{g(\vbeta^{(r)},\bL^{(r)},\vsigma^{2(r)},\vmu^{(r)},\vphi^{(r)};\bv^*)},
\end{equation}
where $(\vbeta^{(1)},\bL^{(1)},\vsigma^{2(1)},\vmu^{(1)},\vphi^{(1)}), \ldots, (\vbeta^{(R_2)},\bL^{(R_2)},\vsigma^{2(R_2)},\vmu^{(R_2)},\vphi^{(R_2)})$ are independent draws from $g(\vbeta,\bL,\vsigma^2,\vmu,\vphi; \bv^*)$ and $p(\by\gvn\vbeta,\bL,\vsigma^2,\vmu,\vphi)$ is the integrated likelihood, which can be estimated using the estimator in \eqref{eq:IS_intlike}. We refer the readers to \citet{CE15} for a more thorough discussion of this adaptive importance sampling approach. We summarize the algorithm in Algorithm~\ref{alg:ml}.

Note that Algorithm~\ref{alg:ml} has two nested importance sampling steps, and it falls within the importance sampling squared (IS$^2$) framework in \citet*{TSPK14}. We follow their recommendation to set $R_1$, the simulation size  of the inner importance sampling loop (the importance sampling step for estimating the integrated likelihood), adaptively so that the variance of the log integrated likelihood is around 1. See also the discussion in \citet*{PdGK12}.

\begin{algorithm}[H]
\caption{Marginal likelihood estimation via the improved cross-entropy method.}
\label{alg:ml}
The marginal likelihood $p(\by)$ can be estimated using the following steps.
\begin{enumerate}
	\item Obtain $M$ posterior draws and use them to solve the CE minimization problem in \eqref{eq:maxMC} to obtain the optimal importance sampling density $g(\vbeta,\bL,\vsigma^2,\vmu,\vphi; \bv^*)$.
		
	\item For $r = 1,\ldots, R_2$, simulate $(\vbeta^{(r)},\bL^{(r)},\vsigma^{2(r)},\vmu^{(r)},\vphi^{(r)}) \sim g(\vbeta^{(r)},\bL^{(r)},\vsigma^{2(r)},\vmu^{(r)},\vphi^{(r)};\bv^*)$ and compute the average
	\[
		\widehat{p(\by)} = \frac{1}{R_2} 	\sum_{r=1}^{R_2}\frac{\widehat{p(}\by\gvn\vbeta^{(r)},\bL^{(r)},\vsigma^{2(r)},\vmu^{(r)},\vphi^{(r)})p(\vbeta^{(r)},\bL^{(r)},\vsigma^{2(r)},\vmu^{(r)},\vphi^{(r)})}
		{g(\vbeta^{(r)},\bL^{(r)},\vsigma^{2(r)},\vmu^{(r)},\vphi^{(r)};\bv^*)},
	\]	
	where the integrated likelihood estimate $\widehat{p(}\by\gvn\vbeta^{(r)},\bL^{(r)},\vsigma^{2(r)},\vmu^{(r)},\vphi^{(r)}) $ is computed using Algorithm \ref{alg:intlike} with $R_1$ independent draws.
\end{enumerate}
\end{algorithm}

\section{Structural Analysis with the VAR-FSV} \label{s:tools}

The VAR-FSV in \eqref{eq:yt}-\eqref{eq:epsilont} can be used to draw structural inference by employing standard tools such as impulse response functions, forecast error variance decompositions and historical decompositions. In particular, letting $\bA(L) = \mathbf{I}_n - \bA_1 L - \dotsm - \bA_p L^p$, where $L$ is the lag operator, the representation
\begin{equation}
\by_t = \vPhi(1)\vmu + \vPhi(L)\vepsilon_t, \label{eq:reduced}
\end{equation}
where $\vPhi(L) = \bA(L)^{-1}$ is well-defined assuming $\det \bA(z) \ne 0$ for all $|z| < 1, z\in \mathbb{C}$ (i.e., the process $\{\by_t : t \in \mathbb{Z}\}$ is covariance-stationary).

Although $\vepsilon_t$ does not contain structural shocks (since its elements are correlated), the reduced-form representation \eqref{eq:reduced} can be matched to a hypothetical structural representation of the form
\begin{equation}
\by_t = \vPhi(1)\vmu + \tilde{\vPhi}_t(L) \bu_t, \label{eq:structural}
\end{equation}
where $\bu_t$ is a vector of structural shock, and hence, its elements are uncorrelated. Note that $\tilde{\vPhi}_t(L)$ is time-varying because $\var(\vepsilon_t) = \bL\vOmega_t\bL' + \vSigma_t$ is time-varying; consequently, hypothetical structural representations that can be matched to \eqref{eq:reduced} will generally have time-varying parameters.

The standard structural VAR approach is to assume that (i) $\bu_t$ is $n \times 1$ and (ii) $\vepsilon_t = \tilde{\vPhi}_{0,t} \bu_t$, where $\tilde{\vPhi}_{0,t}$ is a $n\times n$ constant matrix with $\rank \tilde{\vPhi}_{0,t} = n$ and $\var(\bu_t) = \mathbf{I}_n$. Then, $\tilde{\vPhi}_t(L) = \vPhi(L)\tilde{\vPhi}_{0,t}$ and $\tilde{\vPhi}_{0,t}$ satisfies
\[
\tilde{\vPhi}_{0,t}\tilde{\vPhi}_{0,t}' = \bL\vOmega_t\bL' + \vSigma_t.
\]
In this case, identification of $\tilde{\vPhi}_{0,t}$ requires additional restrictions since
\[
\check{\vPhi}_{0,t}\check{\vPhi}_{0,t}' = \bL\vOmega_t\bL' + \vSigma_t
\]
for all $\check{\vPhi}_{0,t} = \tilde{\vPhi}_{0,t}\bR_t$, given an arbitrary orthonormal matrix $\bR_t$ (i.e. satisfying $\bR_t'\bR_t = \bR_t\bR_t' = \mathbf{I}_n$).

An alternative way to obtain structural inference in our settings---similar to \citet{Korobilis20}---is to assume that $\tilde{\vPhi}_t(L)$ is $n\times(r+n)$ and $\bu_t$ is $(r+n)\times 1$, such that
\begin{align}
\tilde{\vPhi}_{0,t} &= \begin{pmatrix}\bL\vOmega_t^\frac{1}{2} & \vSigma_t^\frac{1}{2}\end{pmatrix}, & \bu_t &= \begin{pmatrix}\tilde{\bff}_t \\ \tilde{\bu}_t^y\end{pmatrix},
\end{align}
where $\tilde{\bff}_t = \vOmega_t^{-\frac{1}{2}}\bff_t$ and $\tilde{\bu}_t^y = \vSigma_t^{-\frac{1}{2}} \bu_t^y$. In this case, $\tilde{\vPhi}_t(L) = \vPhi(L)\tilde{\vPhi}_{0,t}$ is also $n \times (r+n)$, which in departure from standard structural VARs leads to a `short' system \citep{FGS19}.

Identification of impulse response functions and forecast error variance decompositions in short systems is generally problematic \citep{PR22,CF22}. However, in the formulation above, $\tilde{\vPhi}_{0,t}$ is identified due to $\vSigma_t$ being restricted to a diagonal matrix and $\bL$ being identified by sign restrictions as described in Section \ref{s:properties}. Hence, impulse response functions and forecast error variance decompositions to all shocks in $\bu_t$ are identified, even though $\bu_t$ is generally \emph{not recoverable} from past and future observations of $\by_t$, as defined in \citet{CJ21}. 

In our setting, the main interest lies in quantifying the effects of shocks in $\tilde{\bff}_t$, and therefore, sign restrictions on $\bL$ play the role of endowing these shocks with economic meaning. The remaining elements in $\tilde{\bu}_t^y$ are not of direct interest and we do not treat them as economically meaningful shocks. Nevertheless, the restriction that $\vSigma_t$ is diagonal plays a crucial role in the overall identification strategy along with an economically meaningful interpretation of $\tilde{\bff}_t$. We provide explicit expressions for computing impulse response functions and forecast error variance decompositions in Appendix F.

Finally, computing historical decompositions requires $\bff_t$ and $\bu_t^y$ (see Appendix F for details). The fact that $\bu_t$ is not recoverable implies that historical decompositions may not be point identified. In a Bayesian setting, however, the posterior distribution of a historical decomposition at each horizon may still be constructed using draws from the posterior distribution of the VAR-FSV parameters together with draws of $\bff_t$.

In the algorithm developed in Section \ref{s:estimation}, draws of $\bff_t$ are a by-product of simulation, while $\bu_t^y$ is easily obtained as
\[
\bu_t^y = \bA(L)\by_t - \vmu - \bL\bff_t,
\]
for each draw of $\vmu, \bA_1, \dotsc, \bA_p, \bL, \bff_t$. Therefore, draws from the posterior distribution of a HD are straightforward to compute.

In addition, $\bu_t$ can be regarded as being recoverable in the limit as $n\longrightarrow \infty$ from the VAR residual $\vepsilon_t$ (and therefore past and present $\by_t$) under a suitable assumption on the factor loadings $\bL$. To see this, let $\bL^+$ denote the Moore–Penrose inverse of $\bL$. By Assumption \ref{ass2}, $\rank \bL = r$ and $\bL^+ = (\bL'\bL)^{-1}\bL'$. It also follows that a right inverse (although not a Moore-Penrose inverse) of $\tilde{\vPhi}_{0,t}$ is
\begin{align}
\tilde{\vPhi}_{0,t}^{-R} &= \begin{pmatrix} \vOmega_t^{-\frac{1}{2}}\bL^+ \\ \vSigma_t^{-\frac{1}{2}}(\mathbf{I}_n - \bL\bL^+) \end{pmatrix}.
\intertext{Consequently,}
\tilde{\vPhi}_{0,t}^{-R} \tilde{\vPhi}_{0,t} &= \begin{pmatrix} \mathbf{I}_r & \vOmega_t^{-\frac{1}{2}}\bL^+ \vSigma_t^\frac{1}{2} \\ 0 & \mathbf{I}_n - \vSigma_t^{-\frac{1}{2}}\bL\bL^+\vSigma_t^\frac{1}{2} \end{pmatrix}. \label{eq:recov}
\end{align}

In the factor model literature, a standard assumption \citep[e.g.][]{bai2003,FGLR09} is that $n^{-1}\bL'\bL \longrightarrow \vLambda$ as $n \longrightarrow \infty$, where $\vLambda$ is a constant (strictly) positive-definite matrix. It implies that the factors $\bff_t$ are \emph{pervasive} in the sense that they significantly affect most of the variables \emph{on impact}.\footnote{It is worth emphasising, however, that this \emph{does not} imply $\bu_t^y$ is a vector of \emph{idiosyncratic} errors, as defined by \cite{FHLR00,FL01} in the context of generalised dynamic factor models. In particular, the overall effect of $\bu_t^y$ on $\by_t$ is $\vPhi(L)\bu_t^y$, which is generally pervasive, albeit with a delay.} An immediate consequence of the pervasiveness assumption, together the regularity condition that $\var(\epsilon_{i,t}) < \infty$ for all $i = 1, \dotsc, n$, is $\bL^+\vSigma_t^\frac{1}{2}\tilde{\bu}_t^y \overset{m.s.}{\longrightarrow} \mathbf{0}$. Combining this result with \eqref{eq:recov} yields
\begin{equation}
\tilde{\vPhi}_{0,t}^{-R}\vepsilon_t - \bu_t = \left(\tilde{\vPhi}_{0,t}^{-R} \tilde{\vPhi}_{0,t} - \mathbf{I}_n\right)\bu_t \overset{m.s.}{\longrightarrow} \mathbf{0}.
\end{equation}
Consequently, $\bu_t$ is recoverable from $\vepsilon_t$ in the limit.\footnote{A more general result on recoverability with a fixed $n$ is given in \citet{CJ21}.}


\section{A Monte Carlo Study: Determining the Number of Factors} \label{s:MC}

In this section we conduct a series of simulation experiments to assess the adequacy of using the proposed marginal likelihood estimator to determine the number of factors. More specifically, we generate datasets from the VAR-FSV in \eqref{eq:yt}--\eqref{eq:ht2}, but we change the error structure to $\vepsilon_t = \bL \mathbf{f}_t + \sqrt{\theta} \bu_t^y$, where $\theta$ measures the signal-to-noise ratio, following \citet{bai2002determining}. We set parameter values so that if $\theta=r$, the idiosyncratic component will then have the same variance as the common component. In particular, we generate $L_{ij} \sim \mathcal{N}(0, 1)$ for $i=1,\ldots,n$ and $j=1,\ldots,r$ and set $\mu_i=0$ for $i=1,\ldots,n,$ so that the log-volatility  processes associated with the idiosyncratic errors have 0 unconditional mean. 

The remaining parameters are generated as follows. The intercepts are drawn independently from the uniform distribution on the interval $(-10,10)$, i.e., $\distn{U}(-10, 10)$. For the VAR coefficients, the diagonal elements of the first VAR coefficient matrix are iid $\distn{U}(0,0.5)$ and the off-diagonal elements are from $\distn{U}(-0.2,0.2)$; all other elements of the $j$-th ($j > 1$) VAR coefficient matrices are iid $\distn{N}(0,0.1^2/j^2).$ Finally, for the log-volatility processes, we set $\phi_i=0.98$ and $\sigma_i^2 = 0.1^2$ for $i=1,\ldots,n+r$. 

In this Monte Carlo study, we select the true number of factors $r \in \{1,3,5\}$ and $\theta \in \{1,3,5,10\}$; and we consider the number of variables $n \in \{15, 30\}$ and sample size $T \in \{300,500,800\}$. For each set of $(r,\theta, n, T)$, we generate 100 datasets. For each dataset, we estimate the VAR-FSV models with $r=1,\ldots, 6$ factors and compute the associated marginal likelihood values. For this Monte Carlo experiment, a total of 14,400 separate MCMCs and marginal likelihood estimation are run (24 settings $\times$ 6 factor models $\times$ 100 datasets). Among the 6 factor models for each dataset and parameter setting, we select the one with the largest marginal likelihood value. Table~\ref{tab:MCfactor} reports the selection frequency. 

\begin{table}[H]
\centering 
		\caption{Selection frequency (\%) of the number of factors $r$ in 100 datasets.
		The DGP is $\vepsilon_t=\bL\bff_t+\sqrt{\theta} \bu_t^{\by}$. }
\label{tab:MCfactor}
\begin{tabular}{cccccccccc}
\hline \hline
$n$ & $\theta$ & True $r$ & $T$ & $r=1$ & $r=2$ & $r=3$ & $r=4$ & $r=5$ & $r=6$ \\ \hline
15  & 1        & 1   & 300 & 0.90  & 0.10  & 0     & 0     & 0     & 0     \\
\rowcolor{lightgray}
    &          &     & 500 & 0.96  & 0.04  & 0     & 0     & 0     & 0     \\
    &          &     & 800 & 0.99  & 0.01  & 0     & 0     & 0     & 0     \\ 
		\rowcolor{lightgray}
    & 3        & 3   & 300 & 0     & 0.13  & 0.83  & 0.04  & 0     & 0     \\
    &          &     & 500 & 0     & 0.02  & 0.97  & 0.01  & 0     & 0     \\
		\rowcolor{lightgray}
    &          &     & 800 & 0     & 0     & 0.99  & 0.01  & 0     & 0     \\ 
    & 5        & 5   & 300 & 0     & 0.01  & 0.12  & 0.49  & 0.38  & 0     \\
		\rowcolor{lightgray}
    &          &     & 500 & 0     & 0     & 0.01  & 0.25  & 0.74  & 0     \\
    &          &     & 800 & 0     & 0     & 0.01  & 0.05  & 0.94  & 0     \\ 
		\rowcolor{lightgray}
    & 10       & 5   & 300 & 0     & 0.07  & 0.3   & 0.46  & 0.16  & 0.01  \\
    &          &     & 500 & 0     & 0     & 0.1   & 0.48  & 0.42  & 0     \\
		\rowcolor{lightgray}
    &          &     & 800 & 0     & 0     & 0.02  & 0.11  & 0.87  & 0     \\ 
30  & 1        & 1   & 300 & 0.76  & 0.24  & 0     & 0     & 0     & 0     \\
\rowcolor{lightgray}
    &          &     & 500 & 0.97  & 0.02  & 0     & 0     & 0     & 0.01  \\ 
    &          &     & 800 & 1.00  & 0     & 0     & 0     & 0     & 0     \\ 
		\rowcolor{lightgray}
    & 3        & 3   & 300 & 0     & 0.02  & 0.86  & 0.11  & 0.01  & 0     \\
    &          &     & 500 & 0     & 0.01  & 0.98  & 0.01  & 0     & 0     \\
		\rowcolor{lightgray}
    &          &     & 800 & 0     & 0     & 1.00  & 0     & 0     & 0     \\ 
    & 5        & 5   & 300 & 0     & 0     & 0.01  & 0.18  & 0.80  & 0.01  \\
		\rowcolor{lightgray}
    &          &     & 500 & 0     & 0     & 0     & 0.02  & 0.97  & 0.01  \\
    &          &     & 800 & 0     & 0     & 0     & 0.01  & 0.99  & 0     \\ 
		\rowcolor{lightgray}
    & 10       & 5   & 300 & 0     & 0.01  & 0.1   & 0.36  & 0.53  & 0     \\
    &          &     & 500 & 0     & 0     & 0     & 0.16  & 0.84  & 0     \\
		\rowcolor{lightgray}
    &          &     & 800 & 0     & 0     & 0     & 0.02  & 0.98  & 0 \\ \hline \hline
\end{tabular}
\end{table}

The Monte Carlo results show that the marginal likelihood estimator generally performs well in selecting the correct number of factors under a variety of settings. For example, for $n=15,$ $T=300,$ $r = 3$ and $\theta = 3$ (moderate signal-to-noise ratio), the marginal likelihood estimator is able to pick the correct number of factors 83\% of the times; for the rest of the cases, the model with one fewer factor (13\%) or one more factor (4\%) is selected. In addition, as the sample size $T$ increases to 500,  the selection frequency of the $r=3$ factor model increases to 97\%. More generally, the selection frequency of the correct number of factors increases as the sample size $T$ increases for all cases considered, confirming that the marginal likelihood is a consistent model selection criterion. 

\section{Application: The Role of Financial Shocks in Economic Fluctuations} \label{s:application}

We illustrate the proposed methodology by revisiting the structural analysis in \citet{FRS19} that is based on a standard structural VAR. More specifically, they use a 6-variable structural VAR to study the impacts of 5 structural shocks---demand, supply, monetary, investment and financial shocks---on a number of key economic variables, where these structural shocks are identified using sign restrictions on the contemporaneous impact matrix. The size of the VAR in their application is typical among empirical works that use sign restrictions for identification because of the computational burden.\footnote{For their 6-variable structural VAR, \citet{FRS19} report estimation time of about a week using a 12-core workstation.}

However, there are a number of reasons in favor of using a larger set of macroeconomic and financial variables. First, in practice the mapping from variables in an economic model to the data is often not unique. For example, as argued in \citep{LMW21}, the economic variable inflation could be matched to data based on the CPI, PCE, or the GDP deflator, and it is not obvious which time series should be used. Instead of arbitrarily choosing one inflation measure, it is more appropriate to include multiple time series corresponding to the same economic variable in the analysis. 

Second, one might be concerned about the problem of informational deficiency that arises from using a limited information set. More specifically, influential papers such as \citet{HS91} and \citet{LR93, LR94} have pointed out that when the econometrician considers a narrower set of variables than the economic agent, the underlying model used by the econometrician is non-fundamental. That is, current and past observations of the variables do not span the same space spanned by the structural shocks. As a consequence, structural shocks cannot be recovered from the model. A natural way to alleviate this concern of informational deficiency is to use a larger set of relevant variables \citep[see, e.g.,][for a recent review on non-fundamentalness]{gambetti21}.

In view of these considerations, we augment the 6-variable VAR with additional macroeconomic and financial variables, and consider a 20-variable VAR with factor stochastic volatility identified using sign restrictions. There are two related papers that use large VARs to study the role of financial shocks in economic fluctuations. First, \citet{chan21} considers a 15-variable structural VAR with a new asymmetric conjugate prior to identify the financial shocks. Given the larger system and the large number of sign restrictions, estimation time is about a week to obtain 1,000 admissible draws using the algorithm of \citet{RWZ10}. In contrast, the proposed approach takes less than a minute to obtain the same number of admissible draws, and is applicable to even larger systems. Second, \citet{Korobilis20} uses a 15-variable VAR with a factor error structure to identify the financial shocks, which can also be done quickly. The main advantage of our approach, however, is that the structural shocks obtained using our factor stochastic volatility model are point-identified, whereas they are only set-identified under a homoskedastic VAR. In practice, our approach can often provide sharper inference.


\subsection{Data} \label{ss:data}

We use a dataset that consists of 20 US quarterly variables, which are constructed from raw time-series taken from from different sources, including the Federal Reserve Bank of Philadelphia and the FRED database at the Federal Reserve Bank of St. Louis. For easy comparison with the results in \citet{FRS19}, we use the same sample period that spans from 1985:Q1 to 2013:Q2. The complete list of these time-series and their sources are given in Appendix~E. 
 
We include the same 6 variables used in the baseline model in \citet{FRS19}, namely, real GDP, GDP deflator, 3-month treasury rate, ratio of private investment over output, S\&P 500 index and a credit spread defined as the difference between Moody's baa corporate bond yield and the federal funds rate. In addition, we also include 14 additional macroeconomic and financial variables, such as the ratio of total credit over real estate value, labor market variables, mortgage rates, as well as other measures of inflation, interest rates and stock prices. These 20 variables are listed in Table~\ref{tab:sign} and the details of the raw data are given in Appendix~E. 

\subsection{Sign Restrictions and Impulse Responses} \label{ss:IR}

In this section we re-examine the empirical application in \citet{FRS19} that identifies 5 structural shocks using a structural VAR with sign restrictions on the contemporaneous impact matrix. We first use the proposed VAR-FSV model to replicate their baseline results from a 6-variable VAR, but here we impose the sign restrictions on the factor loadings instead of the impact matrix. We then consider a larger VAR-FSV model with 20 variables to identify the same structural shocks. 

Now, we first employ the same 6 variables and the associated sign restrictions used in \citet{FRS19}, which are presented in the first six rows of Table~\ref{tab:sign}. The sign restrictions to identify the supply, demand, monetary, investment and financial shocks are exactly the same as in \citet{FRS19}, and we refer the readers to their paper for more details. Here we only note that in order to distinguish investment and financial shocks from demand shocks, they are assumed to have different effects on the ratio of investment over output. In particular, positive investment and financial shocks have a positive effect on the ratio, motivating by the idea that investment and financial shocks create investment booms. By contrast, positive demand shocks reduce the ratio of investment over output, i.e., even though investment level could increase in response to demand shocks, it does not increase as much as other components of output. 

We compute the impulse responses from the VAR-FSV with 5 factors, where the sign restrictions are imposed on the factor loadings. Since \citet{FRS19} use an improper/non-informative prior in their analysis, to make our results comparable, we consider a proper but relatively vague prior by setting $\kappa_1=\kappa_2=1$.\footnote{The variables are expressed in level. As such, the prior means of the first own lags are all set to be 1, whereas those of other VAR coefficients are set to be 0. In addition, the prior mean of $\mu_i$, the mean of the idiosyncratic log-volatility for the $i$-th variable, is set to be $\log(0.1\times \text{Var}(\mathbf{y}_{i,\cdot}))$. That is, \textit{a priori} about 10\% of the sample variance is attributed to idiosyncratic component. Finally, the prior variance $V_{\mu_{i}}$ is set to be 10 for $i=1,\ldots, n$.} We use the  Gibbs sampler described in Section~\ref{s:estimation} to obtain 50,000 posterior, storing every 10-th draw, after a burn-in period of 5,000.

\begin{table}[H]
\centering 
\caption{Sign restrictions and identified structural shocks.}
\label{tab:sign}
\resizebox{\textwidth}{!}{\begin{tabular}{lccccc}
\hline \hline
                         & Supply & Demand & Monetary & Investment & Financial              \\ \hline
GDP                      & $+$      & +      & +        & +          & +                      \\
\rowcolor{lightgray}
GDP deflator                     & $-$      & +      & +        & +          & +                      \\
3-month tbill rate            & NA     & +      & $-$        & +          & +                      \\
\rowcolor{lightgray}
Investment/output        & NA     & $-$      & NA       & +          & +                      \\
S\&P 500             & +      & NA     & NA       & $-$          & +                      \\
\rowcolor{lightgray}
Spread                   & NA     & NA     & NA       & NA         & NA                     \\ 
Spread 2                   & NA     & NA     & NA       & NA         & NA                     \\
\rowcolor{lightgray}
Credit/Real estate value & NA     & NA     & NA       & NA         & NA \\
Mortgage rates           & NA     & NA     & NA       & NA         & NA \\
\rowcolor{lightgray}
Personal consumption expenditures & +      & +      & +        & +          & +                      \\
Industrial production             & +      & +      & +        & +          & +                      \\
\rowcolor{lightgray}
Industrial production: final      & +      & +      & +        & +          & +                      \\
CPI                 & $-$      & +      & +        & +          & +                      \\
\rowcolor{lightgray}
PCE index           & $-$      & +      & +        & +          & +                      \\
Employment          & NA     & NA     & NA       & NA         & NA \\
\rowcolor{lightgray}
All employees: Manufacturing          & NA     & NA     & NA       & NA         & NA \\
1-year tbill rate                  & NA     & +      & $-$        & +          & +                      \\
\rowcolor{lightgray}
10-year tnote rate  & NA     & +      & $-$        & +          & +                      \\
DJIA                & +      & NA     & NA       & $-$          & +                      \\ 
\rowcolor{lightgray}
NASDAQ        & +      & NA     & NA       & $-$          & +                      \\ \hline
\end{tabular}}
{\raggedright \footnotesize{Note: the variable spread is defined as the difference between Moody's baa corporate bond yield and the federal funds rate. Spread 2 is the difference between Moody's baa corporate bond yield and 10-year treasury yield.} \par}
\end{table}

\begin{figure}[H]
	\begin{center}
		\includegraphics[height=12cm]{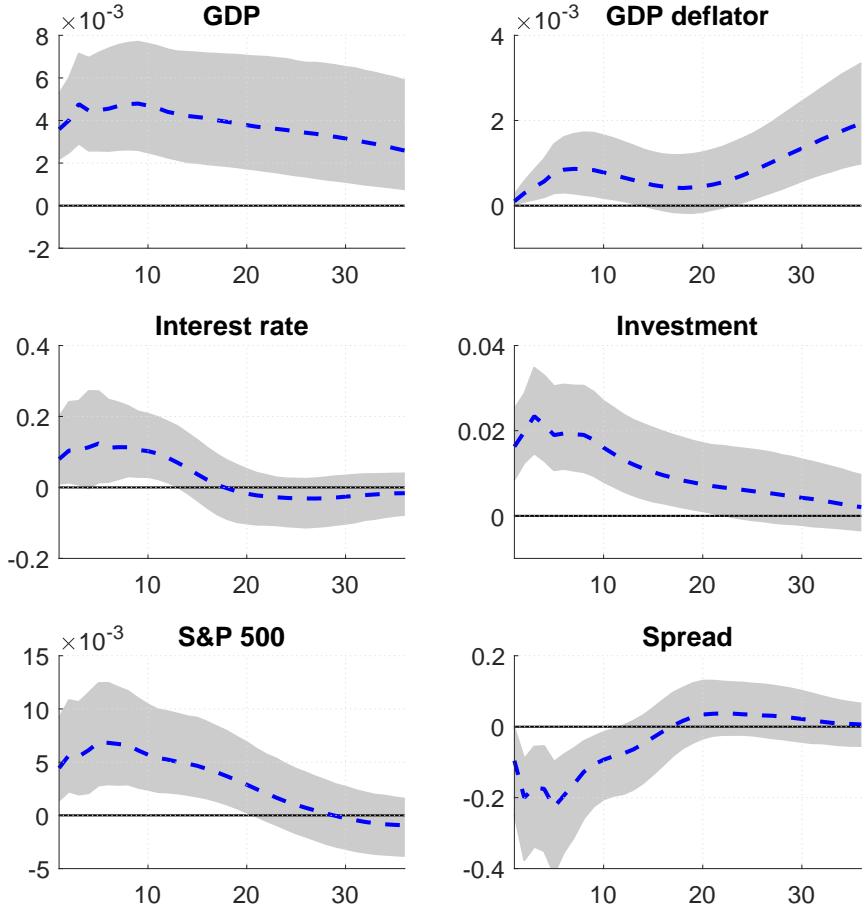}
	\end{center}
	\caption{Impulse responses from a 6-variable VAR-FSV with 5 factors to an one-standard-deviation financial shock. The shaded region represents the 16-th and 84-th percentiles.}
	\label{fig:small-IRF-Financial}
\end{figure}

Figure~\ref{fig:small-IRF-Financial} plots the impulse responses of the 6 variables to an one-standard-deviation financial shock. Despite the differences in methodology, the impulse responses are very similar to those given in \citet{FRS19}. Consistent with the findings in \citet{FRS19}, the results show that financial shocks have a substantial impact on output, stock prices and investment, but have a limited impact on inflation (measured by GDP deflator). Furthermore, even though the impact on the spread is unrestricted, we find that its reaction to financial shocks is significantly counter-cyclical. These results highlight one advantage of the proposed methodology: the median impulse responses from the VAR-FSV are very similar to those obtained using a standard structural VAR, but instead of using an accept-reject algorithm to obtain admissible draws, the sign restrictions can be easily incorporated in the estimation, and consequently, it can be done much faster.
 
Next, we augment the 6-variable VAR with 14 additional macroeconomic and financial variables. Many of these new variables are alternative data series corresponding to the same economic variable. For example, in addition to GDP deflator as prices, we also include CPI and PCE index as alternative measures. Similarly, Dow Jones Industrial Average and NASDAQ indexes are added as alternative measures of stock prices. Furthermore, other seemingly relevant variables, such as labor market and national accounts variables, are also included to alleviate the concern of informational deficiency. The additional variables and the corresponding sign restrictions are listed in rows 7-20 of Table~\ref{tab:sign}.

\begin{figure}[H]
	\begin{center}
		\includegraphics[height=12cm]{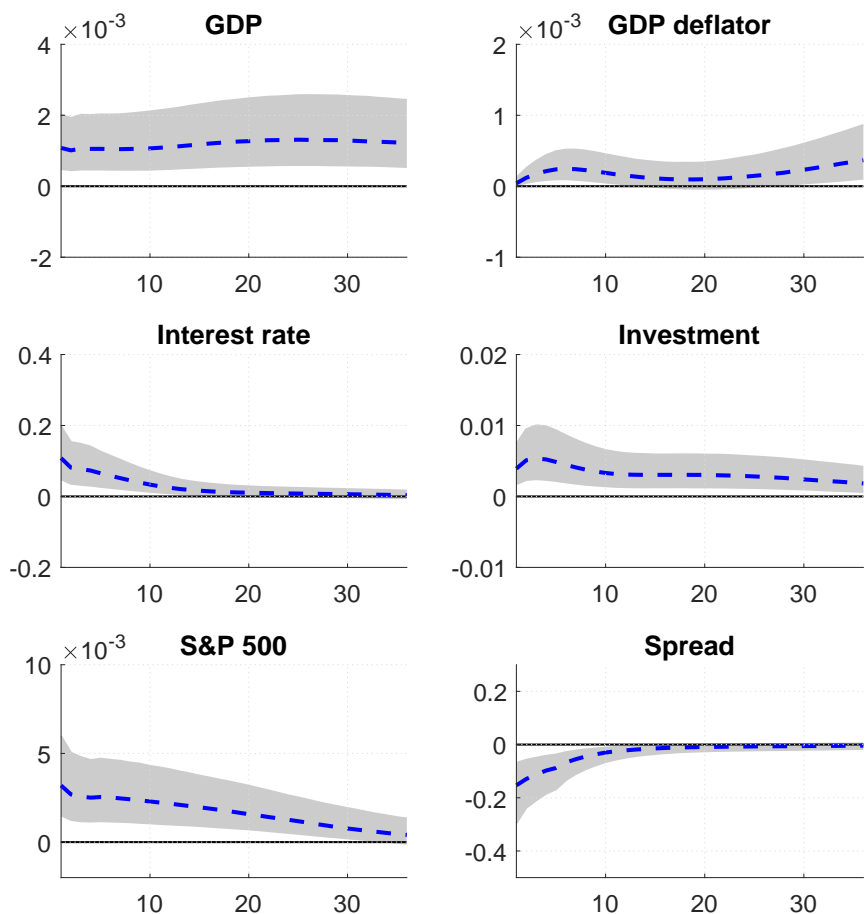}
	\end{center}
	\caption{Impulse responses from a 20-variable VAR-FSV with 5 factors to an one-standard-deviation financial shock. The shaded region represents the 16-th and 84-th percentiles.}
	\label{fig:IRF-Financial}
\end{figure}

With $n=20$ variables and $r=5$ factors, this large VAR-FSV satisfies the condition that $r \leq (n-1)/2$. In addition, it is easy to verify that the sign restrictions given in Table~\ref{tab:sign} satisfy the conditions in Corollary~2, and therefore the latent factors, which we interpret as structural shocks, are point-identified. Given the large number of variables, it is crucial to apply proper shrinkage on the VAR coefficients. Following the bulk of the literature \citep[e.g.,][]{CCM15}, we set $\kappa_1 = 0.04$ and $\kappa_2 = 0.04^2$---i.e., the VAR coefficients associated with lags of other variables are shrunk more strongly to 0 than those on own lags. Again we obtain 50,000 posterior after a burn-in period of 5,000 to compute the impulse responses. The results are reported in Figure~\ref{fig:IRF-Financial}.

The impulse responses from this 20-variable VAR-FSV are qualitatively similar to those from the smaller system, but the inference is much sharper. Specifically, the credible intervals of the 6 impulse response functions are much narrower, highlighting the benefits of incorporating more relevant information---more variables and sign restrictions as well as a more informative prior---to sharpen inference. For example, the credible intervals associated with the responses of investment and stock prices exclude zero for the first 32 quarters after the initial impact of a financial shock. This is in contrast to the much wider credible intervals from the 6-variable VAR (the median impulse responses of stock prices even become negative at longer horizons). The results from this large system therefore better highlight the impact of a positive financial shock, which \citet{FRS19} define as ``a shock that generates an investment and a stock market boom."

\begin{figure}[H]
	\begin{center}
		\includegraphics[height=12cm]{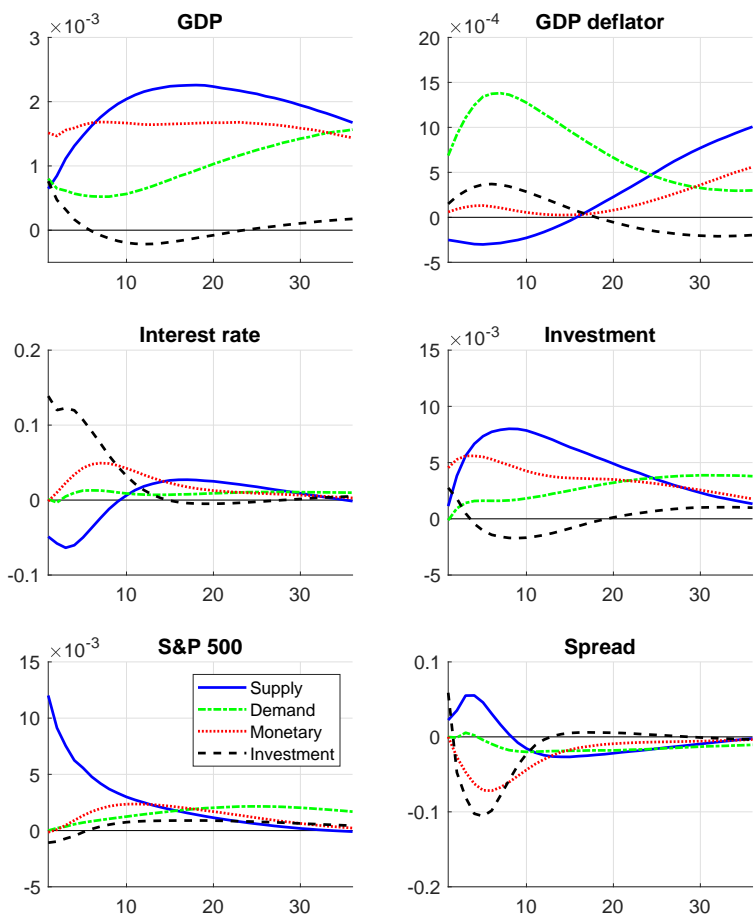} 
	\end{center}
	\caption{Median impulse responses from a 20-variable VAR-FSV with 5 factors to an one-standard-deviation supply, demand, monetary, and investment shock.}
	\label{fig:IRF-other4}	
\end{figure}

Next, we plot in Figure~\ref{fig:IRF-other4} the median impulse responses of the 6 variables from the remaining 4 structural shocks. These impulse responses are similar to those  presented in Figure 3 in \citet{FRS19}. In particular, we confirm that supply shocks generate large effects not only on output, but also on investment and stock prices. On the other hand, demand shocks have smaller effects on output, investment and stock prices, at least for short to medium horizons, but they are the main driver of prices. Finally, while we also find that monetary shocks have a protracted positive effect on output, their effects on stock prices are more subdued. 

\subsection{Historical and Forecast Error Variance Decompositions } \label{ss:decompositions}

To quantify how much of the historical fluctuations in GDP and spread can be attributed to each of the structural shocks, we compute the historical decompositions of these two variables using the formulas derived in Appendix F. The results are reported in Figure~\ref{fig:HD_GDP} and Figure~\ref{fig:HD_spread}.

\begin{figure}[H]
	\begin{center}
		\includegraphics[height=8cm]{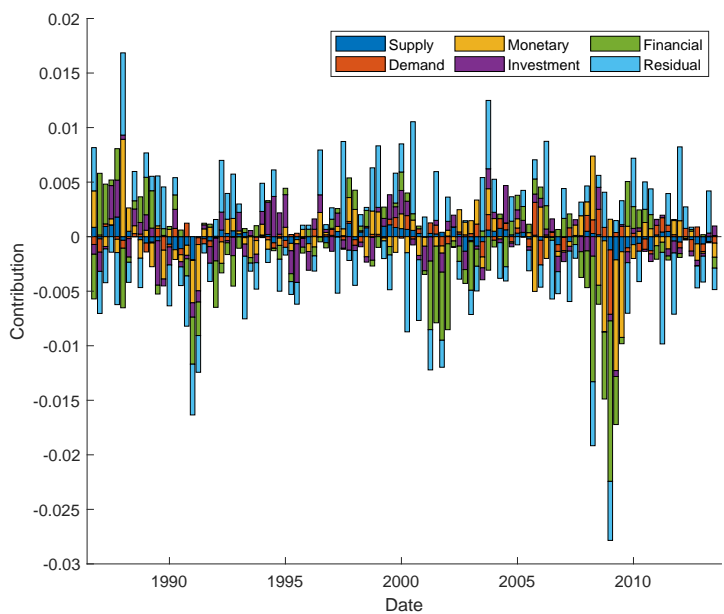} 
	\end{center}
	\caption{Historical decompositions of GDP from  a 20-variable VAR-FSV with 5 factors.}
	\label{fig:HD_GDP}
\end{figure}

\begin{figure}[H]
	\begin{center}
		\includegraphics[height=8cm]{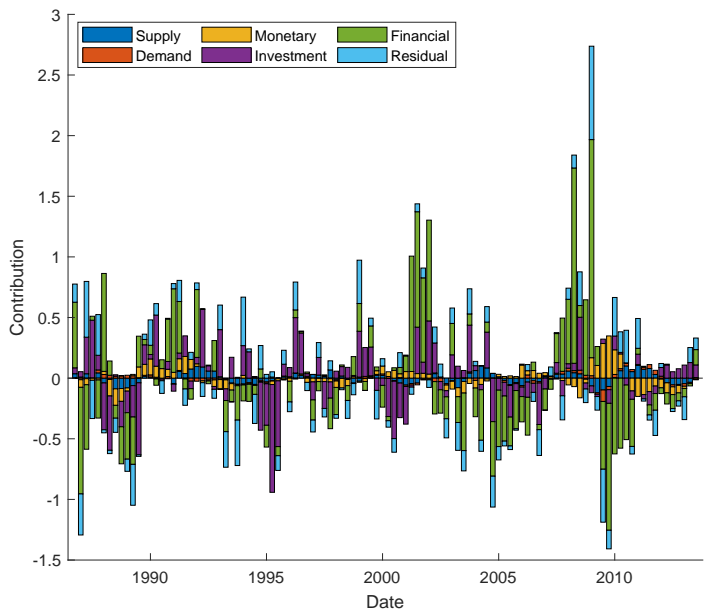} 
	\end{center}
	\caption{Historical decompositions of spread from a 20-variable VAR-FSV with 5 factors.}
	\label{fig:HD_spread}
\end{figure}

These historical fluctuations from the VAR-FSV are in line with those obtained using a standard structural VAR presented in \citet{FRS19}. In particular, financial shocks play a large role in explaining the historical fluctuations in both GDP and spread, especially in the lead-up and aftermath of the Great Recession of 2007-2009.

Next, we quantify the amount of the prediction mean squared errors of 6 selected variables accounted for by each of the 5 structural shocks at different forecast horizons. More specifically, using the expressions developed in Appendix F, we compute the forecast error variance decompositions of the variables and the results are presented in Table~\ref{tab:FEVD}.

\begin{table}[H]
	\footnotesize
	\centering 
	\caption{Forecast error variance decompositions from a 20-variable VAR-FSV with 5 factors (as a percentage of the variation explained by the factors).}
	\label{tab:FEVD}
	\begin{tabular}{llccccc}
		\hline \hline
		& Horizon & Supply & Demand & Monetary & Investment & Financial \\ \hline
		GDP           & 1       & 0.08   & 0.13   & 0.45     & 0.11       & 0.23      \\
		\rowcolor{lightgray}
		& 5       & 0.12   & 0.11   & 0.46     & 0.08       & 0.23      \\
		& 20      & 0.16   & 0.10   & 0.46     & 0.06       & 0.22      \\ 
		\rowcolor{lightgray}
		GDP deflator  & 1       & 0.11   & 0.84   & 0.01     & 0.04       & 0.00      \\
		& 5       & 0.09   & 0.84   & 0.01     & 0.05       & 0.01      \\
		\rowcolor{lightgray}
		& 20      & 0.07   & 0.85   & 0.01     & 0.05       & 0.02      \\ 
		Interest rate & 1       & 0.07   & 0.00   & 0.00     & 0.58       & 0.35      \\
		\rowcolor{lightgray}
		& 5       & 0.10   & 0.00   & 0.00     & 0.58       & 0.32      \\
		& 20      & 0.12   & 0.00   & 0.01     & 0.58       & 0.29      \\ 
		\rowcolor{lightgray}
		Investment    & 1       & 0.07   & 0.01   & 0.44     & 0.16       & 0.33      \\
		& 5       & 0.15   & 0.02   & 0.40     & 0.09       & 0.34      \\
		\rowcolor{lightgray}
		& 20      & 0.23   & 0.02   & 0.37     & 0.06       & 0.32      \\ 
		S\&P 500      & 1       & 0.92   & 0.00   & 0.00     & 0.01       & 0.07      \\
		\rowcolor{lightgray}
		& 5       & 0.91   & 0.00   & 0.00     & 0.01       & 0.07      \\
		& 20      & 0.91   & 0.00   & 0.00     & 0.01       & 0.08      \\ 
		\rowcolor{lightgray}
		Spread        & 1       & 0.02   & 0.00   & 0.00     & 0.13       & 0.85      \\
		& 5       & 0.04   & 0.00   & 0.01     & 0.15       & 0.79      \\
		\rowcolor{lightgray}
		& 20      & 0.07   & 0.01   & 0.04     & 0.19       & 0.69         \\ \hline
	\end{tabular}
\end{table}

Overall, financial shocks play a large role in explaining the forecast error variances of the majority of the variables. The two exceptions are prices (measured by GDP deflator), which are mainly impacted by demand shocks, and stock prices (measured by S\&P 500 index), which are mostly driven by supply shocks.

\section{Concluding Remarks} \label{s:conclusion}

We have considered an order-invariant VAR with factor stochastic volatility and shown how the presence of multivariate stochastic volatility allows for statistical identification of the model. Furthermore, we have worked out sufficient conditions in terms of sign restrictions on the impact of the structural shocks for point-identification of the corresponding structural model. To estimate the proposed order-invariant VAR, we developed an efficient MCMC algorithm that can incorporate a large number of variables and sign restrictions. In an empirical application involving 20 macroeconomic and financial variables, we demonstrated the ability of our methods to produce more precise impulse responses compared to a medium-sized structural VAR.

\newpage

\section*{Appendix A: Proofs of Propositions}

In this appendix we provide the proofs of the propositions and corollaries stated in the main text. To that end, we first consider the following two lemmas.

\begin{lemma}[\citet{magnus2019matrix}, Theorem 2.13, pp. 43] \rm
A necessary and sufficient condition for the matrix equation $\bA\bX\bB = \bC$ to have a solution is that
\begin{equation}
    \bA\bA^+\bC\bB^+\bB = \bC, \label{MNeq1}
\end{equation}
where $\bD^+$ denotes the Moore–Penrose inverse of $\bD$. In this case the general solution is 
\begin{equation}
    \bX = \bA^+\bC\bB^+ +\bQ-\bA^+\bA\bQ\bB\bB^+ \label{MNeq2}
\end{equation}
where $\bQ$ is an arbitrary matrix of the appropriate dimension. In particular, if $\bA$ has full column rank and $\bB$ has full row rank, then the unique solution is given by:
\begin{equation}
    \bX = \bA^+\bC\bB^+.  \label{MNeq3}
\end{equation}
\end{lemma}

\textbf{Proof of Lemma 1}: The proof that \eqref{MNeq1} is the necessary and sufficient condition and that the general solution has the form in \eqref{MNeq2} follows directly from \citet{magnus2019matrix}. For uniqueness \eqref{MNeq3}, note that if $\bA$ and $\bB$ have full column and row rank, respectively, their Moore–Penrose inverses can be computed as:
\[
	\bA^+ = (\bA'\bA)^{-1}\bA', \quad \bB^+ = \bB'(\bB\bB')^{-1}.
\]
It then follows that $\bA^+\bA=\mathbf{I}$ and $\bB \bB^+=\mathbf{I}$, and therefore \eqref{MNeq2} reduces to \eqref{MNeq3}. $\qed$

The next lemma adapts a theorem in \citet{AR56} to our setting with heteroskedastic factors. 

\begin{lemma}[\citet{AR56}, Theorem 5.1, pp. 118] \label{lemma2} \rm Under Assumption~\ref{ass2}, the common and idiosyncratic variance components are separately identified. That is, for two observationally equivalent models such that $\vGamma_t =\bL\vOmega_t\bL'+ \vSigma_t =  {\bL}^*{\vOmega}^*_t{\bL}^{*\prime}+ {\vSigma}^*_t$, we have $\bL^*\vOmega^*_t\bL^{*\prime} =\bL\vOmega_t\bL'$ and $\vSigma_t^* = \vSigma_t$.
\end{lemma}

\noindent \textbf{Proof of Lemma~2}: Suppose we have two observationally equivalent models such that $\vGamma_t =\bL\vOmega_t\bL'+ \vSigma_t =  {\bL}^*{\vOmega}^*_t{\bL}^{*\prime}+ {\vSigma}^*_t$. We wish to show that $\bL^*\vOmega^*_t\bL^{*\prime} =\bL\vOmega_t\bL'$ and $\vSigma_t^* = \vSigma_t$. Since the
off-diagonal elements of $\bL\vOmega_t\bL'$ and of $\bL^*\vOmega^*_t\bL^{*\prime}$ are the corresponding off-diagonal elements of $\vGamma_t$, it suffices to show that the diagonal elements of $\bL\vOmega_t\bL'$ are equal to the diagonal elements of $\bL^*\vOmega^*_t\bL^{*\prime}$. First note that Assumption~\ref{ass2} implies that $2r +1 \leq n$. Furthermore, let
\[
    \bL=\begin{pmatrix}
\bL_1   \\
l_{r+1} \\
\bL_2   \\
\bL_3 \\
\end{pmatrix} ,  \ \ \ 
    \bL^*=\begin{pmatrix}
\bL_1^*   \\
l_{r+1}^* \\
\bL_2^*   \\
\bL_3^* \\
\end{pmatrix} 
\]
where $\bL_1$ and $\bL_2$ are nonsingular square matrices of dimension $r\times r$, $l_{r+1}$ is the 
$(r+1)$-th row, and $\bL_3$ is of dimension $(n-2r-1)\times r$ (it can be null if $n=2r+1$); $\bL^*$ is partitioned into submatrices similarly.
Then, we have
\[
    \bL\vOmega_t\bL'=\begin{pmatrix}
\bL_1\vOmega_t\bL_1'  & \bL_1 \vOmega_t l_{r+1}'  &  \bL_1 \vOmega_t  \bL_2' & \bL_1  \vOmega_t \bL_3' \\
l_{r+1} \vOmega_t \bL_1'  & l_{r+1} \vOmega_t  l_{r+1}' & l_{r+1} \vOmega_t  \bL_2' & l_{r+1}  \vOmega_t \bL_3' \\
\bL_2 \vOmega_t \bL_1'   & \bL_2 \vOmega_t  l_{r+1}' & \bL_2  \vOmega_t \bL_2' &  \bL_2 \vOmega_t  \bL_3' \\
\bL_3 \vOmega_t \bL_1'   & \bL_3  \vOmega_t l_{r+1}' &  \bL_3 \vOmega_t  \bL_2' & \bL_3 \vOmega_t  \bL_3' \\
\end{pmatrix} 
\]
and $\bL^*\vOmega^*_t\bL^{*\prime}$ has the same form. Since $\bL_1 \vOmega_t l_{r+1}'$, $ \bL_1 \vOmega_t  \bL_2' $, $l_{r+1} \vOmega_t  \bL_2'$ are off-diagonal, $\bL_1 \vOmega_t l_{r+1}'=\bL_1^* \vOmega_t^* l_{r+1}^{*\prime}$, $ \bL_1 \vOmega_t  \bL_2'=\bL_1^* \vOmega_t^*  \bL_2^{*\prime} $ and $l_{r+1} \vOmega_t  \bL_2'=l_{r+1}^* \vOmega_t^*  \bL_2^{*\prime}$. Note that since $\bL_1$ and $\bL_2$ are nonsingular, so is 
 $\bL_1 \vOmega_t  \bL_2'$. Next, since $\bL\vOmega_t\bL'$ is of rank $r$, any square submatrix of dimension larger than $r$ is singular. In particular, 
\begin{align*} 
    0&=\begin{vmatrix}
\bL_1^* \vOmega_t^* l_{r+1}^{*\prime}  &  \bL_1^* \vOmega_t^*  \bL_2^{*\prime}\\
l_{r+1}^* \vOmega_t^*  l_{r+1}^{*\prime} & l_{r+1}^* \vOmega_t^*  \bL_2^{*\prime} \\
\end{vmatrix} =
\begin{vmatrix}
\bL_1 \vOmega_t l_{r+1}'  &  \bL_1 \vOmega_t  \bL_2'\\
l_{r+1}^* \vOmega_t^*  l_{r+1}^{*\prime} & l_{r+1} \vOmega_t  \bL_2' \\
\end{vmatrix} \\
&=(-1)^r l_{r+1}^* \vOmega_t^*  l_{r+1}^{*\prime} |\bL_1 \vOmega_t  \bL_2'| + f(\bL\vOmega_t\bL').
\end{align*}
Similarly, $0 = (-1)^r l_{r+1} \vOmega_t  l_{r+1}^{\prime} |\bL_1 \vOmega_t  \bL_2'| + f(\bL\vOmega_t\bL')$. Since $|\bL_1 \vOmega_t  \bL_2'| \neq 0$, we must have $l_{r+1} \vOmega_t  l_{r+1}^{\prime}=l_{r+1}^* \vOmega_t^*  l_{r+1}^{*\prime}$. In the same fashion, we can show that the other diagonal elements of 
$\bL\vOmega_t\bL'$ are equal to those of $\bL^*\vOmega^*_t\bL^{*\prime}$. $\qed$

\noindent \textbf{Proof of Proposition 2}: Suppose we have two observationally equivalent models such that $\vGamma_t =\bL\vOmega_t\bL'+ \vSigma_t=  {\bL}^*{\vOmega}^*_t{\bL}^{*\prime}+ {\vSigma}^*_t$. Under Assumption~\ref{ass2}, Lemma~\ref{lemma2} implies that the common and the idiosyncratic variance components can be separately identified, i.e., $\bL^*\vOmega^*_t\bL^{*\prime} =\bL\vOmega_t\bL'$ and $\vSigma_t^* = \vSigma_t$. 

For notational convenience, let $\vect{\omega}_t=\text{vecd}(\vOmega_t)=(\omega_{1,t}, \ldots, \omega_{r,t})'$ and $\vect{\sigma}_t=\text{vecd}(\vSigma_t)=(\sigma_{1,t}, \ldots, \sigma_{n,t})'$. Consider the first identity $\bL\vOmega_t\bL'= \bL^*\vOmega^*_t\bL^{*\prime}$. By Lemma 1, the necessary and sufficient condition to solve the system of equations for $\vOmega_t^*$ is 
\begin{equation}
   \bL^*\bD^*\vOmega_t\bD^{*\prime}\bL^{*\prime}- \bL \vOmega_t\bL'=\vect{0} \label{id1-eq1}
\end{equation}
where $\bD^* = [\bL^{*}]^{+}\bL$ and $[\bL^{*}]^{+}=(\bL^{*\prime}\bL^*)^{-1}\bL^{*\prime}$ since $\bL^*$ has full column rank. Let $l_{ij}$ and $l^d_{ij}$ denote, respectively, the $(i,j)$ element of $\bL$ and the product $\bL^*\bD^*$. Equation~\eqref{id1-eq1} implies that $\sum_{k=1}^r\omega_{k,t}(l^d_{ik}l^d_{jk}-l_{ik}l_{jk})=0$, $i, j=1,\ldots, n$. Under the assumption that the elements in $\vect{\omega}_t$ are linearly independent (i.e., the only solution to $\vdelta'\vect{\omega}_t = 0 $ for all $t$ is $\vdelta=\vect{0}$), we must have $\bL^*\bD^*={\pm}\bL$. Hence, $\bL^*$ is obtained once $\bD^*$ is determined. We consider $\bL^*\bD^*=\bL$. As will be clear later, the same conclusion applies for the case $\bL^*\bD^*=-\bL$.

We next turn to the determination of $\bD^*$. Since $\bL^*$ has full column rank, again by Lemma~1, the unique solution to $\bL^*\vOmega^*_t\bL^{*\prime} = \bL\vOmega_t\bL'$ is 
$\vOmega^*_t = \bD^*\vOmega_t\bD^{*\prime}$. In particular, since $\vOmega_t^*$ is diagonal, we have $\sum_{l=1}^r\omega_{l,t}d^*_{il}d^*_{jl}=0$ for $i,j=1,\ldots, r, i\neq j$ and $t=1,\ldots,T$. These restrictions can be more succinctly expressed as $\dot{\vOmega}_T \mathbf{b}_{ij} = \mathbf{0}$, where $\mathbf{b}_{ij}=(d^*_{i1}d^*_{j1},\ldots,d^*_{ir}d^*_{jr})'$ and $\dot{\vOmega}_T = (\vect{\omega}_{1},\ldots,\vect{\omega}_{T})'$ is $T\times r$. Given that the rank of $\dot{\vOmega}_T$ is $r$ when the processes in $\vect{\omega}_t$ are linearly independent for $t=1,\ldots, T$, the only solution to such a set of $T$ homogeneous linear equations is $\mathbf{b}_{ij} = \mathbf{0}$ irrespective of $i$ and $j$. Therefore, each column of $\bD^*$ contains at most one nonzero element (otherwise for some column $k$ there exist nonzero elements $d_{ik}$ and $d_{jk}$ with $i\neq j$ such that $d_{ik} d_{jk} \neq 0$, contradicting $\mathbf{b}_{ij} = \mathbf{0}$). In this scenario, similar to \citet{BB20}, we can write $\bD^*=\bP_{\pm}\bP_{r}\bS_{r}$, where $\bP_{r}$ is one of the $r!$ permutation matrices, $\bP_{\pm} = \diag(\pm 1, \ldots, \pm 1)$ is a reflection matrix that corresponds to one of the $2^r$ ways to switch the signs of the $r$ columns, and $\bS_{r}$ is an arbitrary diagonal scaling matrix of dimension $r\times r$. 

Next, we show  that $\bS_{r}$ must be an identity matrix. Using the fact that 
$\vOmega^*_t = \bD^*{\vOmega}_t\bD^{*\prime}$, we can write the observationally equivalent factors as $\mathbf{f}^*_t=\bD^*\mathbf{f}_t$. Without loss of generality, we consider solely the scaling effect, i.e., $\bD^*=\bS_{r} = \diag(s_1,\ldots,s_r)$. Now, $\mathbf{f}^*_t \sim\distn{N}(\mathbf{0}, {\vOmega}_t^*)$, where
\begin{align}
	\vOmega_t^* & =\bS_{r}\text{diag}(\e^{h_{n+1,t}},\ldots, \e^{h_{n+r,t}})\bS_{r}^{'} \notag \\
	& = \text{diag}(\e^{h_{n+1,t}+2 \log s_1 },\ldots, \e^{h_{n+r,t}+2\log s_r}) \notag \\
	& \equiv \text{diag}(\e^{h^*_{n+1,t}},\ldots, \e^{h^*_{n+r,t}}). \label{standardization}
\end{align}
Since we standardize the unconditional variances of the log stochastic volatility processes to be one, we must have $\Em h^*_{n+j,t} = \Em [h_{n+j,t}+2\log s_j] = 0+2\log s_j =0$ for $j=1,\ldots,r$, which implies that $s_j = 1$. Thus, $\bS_{r} = \mathbf{I}_r$, and the only form $\bD^*$ can take is $\bD^*=\bP_{\pm}\bP_{r}$. \qed

\noindent \textbf{Proof of Proposition 3}: The proposition is equivalent to the claim that the only feasible factor loadings submatrix $\bL_1^*$ under the assumptions must satisfy $\bL_1^*=\bL_1 \bD_1$, where $\bD_1=\bP_{\pm}\bP_{r_1}$. In what follows, we prove the claim by using a similar approach as in the proof of Proposition 2. First notice that since $\bL$ satisfies Assumption~\ref{ass2}, any observationally equivalent model must satisfy ${\bL}^*{\vOmega}^*_t{\bL}^{*\prime}=\bL\vOmega_t\bL'$. Applying the same argument as in the proof of Proposition 2, the solution ${\bL}^*$ must be in the form ${\bL}={\bL}^*\bD^*$, and it follows that $\vOmega^*_t=\bD^*{\vOmega}_t\bD^{*\prime}$. Partition $\bD^*$ conformably as $\bD^*=(\bD^*_1, \bD^*_2)$, where $\bD^*_1$ is $r\times r_1$ and $\bD^*_2$ is $r\times r_2$. We thus have
\begin{equation} \label{part1}
    \vOmega^*_t=(\bD^*_1, \bD^*_2) \begin{pmatrix}
\vOmega_{1t} & \mathbf{0}  \\
\mathbf{0} & \mathbf{I}_{r_2}
\end{pmatrix}\begin{pmatrix}
\bD^{*\prime}_1 \\
\bD^{*\prime}_2 
\end{pmatrix}=\bD^*_1 \vOmega_{1t} \bD^{*\prime}_1+\bD^*_2 \bD^{*\prime}_2.
\end{equation}
Again, since ${\vOmega}_t^*$ is diagonal, the off-diagonal elements of $\bD^*_1 \vOmega_{1t} \bD^{*\prime}_1+\bD^*_2  \bD^{*\prime}_2$ must be zero, i.e., $\sum_{l_1=1}^{r_1}{\omega}_{l_1,t}d^*_{il_1}d^*_{jl_1}+\sum_{l_2=r_1+1}^{r}d^*_{il_2}d^*_{jl_2}=0$ for $j>i$ and $t=1,\ldots,T$, where ${{\omega}}_{l_1,t}$ is the $l_1$-th element in ${\vect{\omega}}_{t}=({{\omega}}_{1,t},\ldots,{{\omega}}_{r_1,t},1,\ldots,1)'$. 

For a given pair $(i,j)$, these restrictions can be expressed as $\ddot{\vOmega}_t \mathbf{b}_{ij}=\mathbf{0}_T$, where $\ddot{\vOmega}_t=(\dot{\vect{\omega}}_{1},\ldots,{\dot{\vect{\omega}}}_{t})'$ is a $T\times (r_1+1)$ matrix, $\dot{\vect{\omega}}_{t}=({{\omega}}_{1,t},\ldots,{{\omega}}_{r_1,t},1)'$ for $t=1,\ldots,T$, $\mathbf{b}_{ij}=(d^*_{i1}d^*_{j1},\ldots,d^*_{ir_1}d^*_{jr_1},\sum_{l_2=r_1+1}^{r}d^*_{il_2}d^*_{jl_2})'$. Given that the rank of $\ddot{\vOmega}_t$ is $r_1+1$ when the processes in $\dot{\vect{\omega}}_t$ are linearly independent, the only solution to such a set of $T$ homogeneous linear equations is $\mathbf{b}_{ij}=\mathbf{0}_{r_1}$ irrespective of $i$ and $j$. Therefore, applying the same argument as before, the condition that the first $r_1$ restrictions in $\mathbf{b}_{ij}=\mathbf{0}_{r_1}$---i.e., $d^*_{i1}d^*_{j1}= \cdots =d^*_{ir_1}d^*_{jr_1}=0$---implies that each column of $\bD_1^*$ contains at most one element that is different from 0. 

Next, we show that any nonzero elements can only be in the upper $r_1$ rows of $\bD_1^*$, which in turn makes the lower $r_2$ rows a zero submatrix. To that end, we partition $\bD^*$ conformably and write \eqref{part1} as:
\begin{align*} 
    \vOmega^*_t & =
\begin{pmatrix}
\bD_{11}^* & \bD_{12}^*  \\
\bD_{21}^* & \bD_{22}^*
\end{pmatrix} \begin{pmatrix}
\vOmega_{1t} & \mathbf{0}  \\
\mathbf{0} & \mathbf{I}_{r_2}
\end{pmatrix}\begin{pmatrix}
\bD_{11}^{*\prime} & \bD_{21}^{*\prime}  \\
\bD_{12}^{*\prime} & \bD_{22}^{*\prime}
\end{pmatrix}  \\
& =\begin{pmatrix}
\bD_{11}^*\vOmega_{1t}\bD_{11}^{*\prime}+\bD_{12}^*\bD_{12}^{*\prime} & \bD_{11}^*\vOmega_{1t}\bD_{21}^{*\prime}+\bD_{12}^*\bD_{22}^{*\prime}  \\
\bD_{21}^*\vOmega_{1t}\bD_{11}^{*\prime}+\bD_{22}^*\bD_{12}^{*\prime} & \bD_{21}^*\vOmega_{1t}\bD_{21}^{*\prime}+\bD_{22}^*\bD_{22}^{*\prime} 
\end{pmatrix}. 
\end{align*}
Since $\vOmega^*_t$ is diagonal, we must have
\begin{align}
 \bD_{21}^*\vOmega_{1t}\bD_{21}^{*\prime}+\bD_{22}^*\bD_{22}^{*\prime} &= \mathbf{I}_{r_2} \label{eq0:1} \\
	\bD_{11}^*\vOmega_{1t}\bD_{21}^{*\prime}+\bD_{12}^*\bD_{22}^{*\prime} &= \mathbf{0} \label{eq0:2} \\
	\bD_{21}^*\vOmega_{1t}\bD_{11}^{*\prime}+\bD_{22}^*\bD_{12}^{*\prime} &= \mathbf{0} \notag 	
\end{align}
and $\bD_{11}^*\vOmega_{1t}\bD_{11}^{*\prime}+\bD_{12}^*\bD_{12}^{*\prime}$ is diagonal. Using exactly the same argument in analyzing \eqref{part1}, $\bD_{11}^*\vOmega_{1t}\bD_{11}^{*\prime}+\bD_{12}^*\bD_{12}^{*\prime}$ being diagonal implies a set of $T$ homogeneous linear equations of dimension $r_1+1$. It follows that each column of $\bD_{11}^*$ contains at most one nonzero element. Since earlier we have proved the same result for $\bD^*_1=(\bD^{*\prime}_{11},\bD^{*\prime}_{21})'$, it must be the case that all the nonzero elements are in $\bD^{*}_{11}$, i.e., $\bD^{*}_{21}=\mathbf{0}$. Otherwise, there is a least one row in $\bD_{11}^*$ whose elements are all 0, say row $k$ with $k\leq r_1$, which implies that 
\begin{align}
[\vOmega^*_t]_{(k,k)}&= [\bD_{11}^*\vOmega_{1t}\bD_{11}^{*\prime}]_{(k,k)}+[\bD_{12}^*\bD_{12}^{*\prime}]_{(k,k)} \notag \\
&=\sum_{l_1=1}^{r_1}{\omega}_{l_1,t}d^*_{kl_1}d^*_{kl_1} +  \sum_{l_2=r_1+1}^{r}d^{*2}_{kl_2}=0+\sum_{l_2=r_1+1}^{r}d^{*2}_{kl_2} = \text{constant}, \label{one}
\end{align}
where $[\bA]_{(i,j)}$ denotes the $(i,j)$-th element of $\bA$. It is clear that \eqref{one} violates  the assumption that  $(\text{vecd}(\vOmega_{1t}^*)', 1)'$ are linearly independent for all $t$.

Now, using the fact that $\bD^{*}_{21}=\mathbf{0}$ in \eqref{eq0:1}, it follows $\bD_{22}^*$ is an orthogonal matrix. Next, using the fact that $\bD^{*}_{21}=\mathbf{0}$ in \eqref{eq0:2}, we have $\bD^*_{12}=\mathbf{0}$ since the orthogonal matrix $\bD^*_{22}$ is invertible. Subsequently, the $(1,1)$-th block of $\vOmega_t^*$ reduces to $\bD_{11}^*\vOmega_{1t}\bD_{11}^{*\prime}$. From the earlier conclusion that each column of $\bD_{11}^*$ has at most one nonzero element and  the standardization requirement as shown in \eqref{standardization}, it is clear that $\bD_{11}^*$ must be of the form $\bP_{\pm}\bP_{r_1}$. To summarize, we have shown that
\[
   \bL^*=(\bL^*_1, \bL^*_2) = \bL \bD^{*\prime}= (\bL_1, \bL_2)\begin{pmatrix}
\bP_{\pm}\bP_{r_1} & \mathbf{0}  \\
\mathbf{0}  & \bD_{22}^{*\prime}
\end{pmatrix} = (\bL_1\bP_{\pm}\bP_{r_1}, \bL_2\bD_{22}^{*\prime}),
\]
where $\bD_{22}^{*}$ is an orthogonal matrix of dimension $r_2$. $\qed$

\noindent \textbf{Proof of Corollary 1}: The proof follows directly from the proof of Proposition 3. More specifically, under the assumption  $r_1=r-1$, $\bD_{22}^{*}$ is an orthogonal matrix of dimension~1. Thus the only admissible $\bD_{22}^{*}$ is $\pm 1$. So the full matrix $\bD^{*}$ is also of the form $\bP_{\pm}\bP_{r}$. $\qed$

\newpage

\section*{Appendix B: Estimation Details}

In this appendix we provide the estimation details for fitting the model in~\eqref{eq:yt}--\eqref{eq:ht2}. More specifically, posterior draws can be obtained by sampling sequentially from the following distributions: 
\begin{enumerate}
	\item $p(\mathbf{f} \gvn \by, \vbeta, \bL, \bh, \vmu, \vphi,\vsigma^2) 
	= p(\mathbf{f} \gvn \by, \vbeta, \bL, \bh)$; 
	
	\item $p(\vbeta,\bL \gvn \by, \mathbf{f}, \bh, \vmu, \vphi,\vsigma^2) = 
	\prod_{i=1}^n p(\vbeta_i,\bl_i \gvn \by_{i,\bigcdot}, \mathbf{f}, \bh_{i,\bigcdot})$; 
	
	\item $p(\bh \gvn \by, \mathbf{f},\vbeta, \bL, \vmu, \vphi,\vsigma^2) = 
	\prod_{i=1}^{n+r} p(\bh_{i,\bigcdot} \gvn \by, \mathbf{f}, \vbeta, \bL, \vmu, \vphi,\vsigma^2)$; 
	
	\item $p(\vsigma^2 \gvn \by, \mathbf{f}, \vbeta, \bL, \bh, \vmu, \vphi) 	
	= \prod_{i=1}^{n+r} p(\sigma_i^2 \gvn \bh_{i,\bigcdot}, \mu_i, \phi_i)$;
	
	\item $p(\vmu \gvn \by, \mathbf{f}, \vbeta, \bL, \bh, \vphi,\vsigma^2) 
	= \prod_{i=1}^{n} p(\mu_i \gvn \bh_{i,\bigcdot}, \phi_i, \sigma^2_i) $; 
	
	\item $p(\vphi \gvn \by, \mathbf{f}, \vbeta, \bL, \bh, \vmu,\vsigma^2) 
	= \prod_{i=1}^{n+r} p(\phi_i \gvn \bh_{i,\bigcdot}, \mu_i, \sigma^2_i) $.	
	
\end{enumerate}

In Section~\ref{s:estimation} of the main text we describe the implementation details of Step 1 and Step2. Below we give the details of the remaining steps. 

\textbf{Step 3}: Sample $\bh$. Again given the latent factors $\mathbf{f}$, the VAR becomes $n$ unrelated regressions and we can sample each vector of log-volatilities $\bh_{i,\bigcdot} = (h_{i,1},\ldots, h_{i,T})'$ separately. More specifically, we can directly apply the auxiliary mixture sampler in \citet*{KSC98}  in conjunction with the precision sampler of \citet{CJ09} to sample from $(\bh_{i,\bigcdot} \gvn \by, \mathbf{f},\vbeta,\bL,\vmu,\vphi,\vsigma^2)$ for $i=1,\ldots, n+r$. For a textbook treatment, see, e.g., \citet{CKPT19} chapter 19.

\textbf{Step 4}: Sample $\vsigma^2$. This step can be done easily, as the elements of $\vsigma^2$ are conditionally independent and, for $i=1,\ldots, n+r$, each follows an inverse-gamma distribution:
\[
	(\sigma_i^2 \gvn \bh_{i,\bigcdot},\mu_i,\phi_i) \sim \distn{IG}(\nu_{i}+T/2, \widetilde{S}_i),
\]
where $ \widetilde{S}_i = S_i + [(1-\phi_i^2)(h_{i,1}-\mu_i)^2 + \sum_{t=2}^T(h_{i,t}-\mu_i-\phi_i(h_{i,t-1}-\mu_i))^2]/2$, with the understanding that $\mu_i=0$ for $i>n$.

\textbf{Step 5}: Sample $\vmu$. It is also straightforward to implement this step, as the elements of $\vmu$ are conditionally independent and, for $i=1,\ldots, n$, each follows a normal distribution:
\[
	(\mu_i \gvn \bh_{i,\bigcdot}, \phi_i, \sigma^2_i)\sim \distn{N}(\hat{\mu}_i, K_{\mu_i}^{-1}),
\]
where
\begin{align*}
	K_{\mu_i} & = V_{\mu_i}^{-1} + \frac{1}{\sigma_i^2}\left[ 1-\phi_i^2 + (T-1)(1-\phi_i)^2\right] \\
	\hat{\mu}_i & = K_{\mu_i}^{-1}\left[V_{\mu_i}^{-1}\mu_{0,i} + \frac{1}{\sigma_i^2}\left( (1-\phi_i^2)h_{i,1} + (1-\phi_i)\sum_{t=2}^T(h_{i,t}-\phi_ih_{i,t-1})\right)\right].
\end{align*}

\textbf{Step 6}: To sample $\phi_i, i=1,\ldots, n+r$, note that 
\[
	p(\phi_i \gvn \bh_{i,\bigcdot},\mu_i,\sigma^2_i)\propto p(\phi_i)g(\phi_i)\e^{-\frac{1}{2\sigma^2_i}\sum_{t=2}^T(h_{i,t}-\mu_i-\phi_i(h_{i,t-1}-\mu_i))^2},
\]
where $ g(\phi_i) = (1-\phi_i^2)^{\frac{1}{2}}\e^{-\frac{1}{2\sigma_i^2}(1-\phi_i^2)(h_{i,1}-\mu_i)^2}$ and $p(\phi_i)$ is the truncated normal prior, with the understanding that $\mu_i=0$ for $i>n$. The conditional density $p(\phi_i \gvn \bh_{i,\bigcdot},\mu_i,\sigma^2_i)$ is non-standard, but a draw from it can be obtained by using 
an independence-chain Metropolis-Hastings step with proposal distribution $\distn{N}(\hat{\phi}_i, K_{\phi_i}^{-1}) 1(|\phi_i|<1)$, where 
\begin{align*}
	K_{\phi_i}   & = V_{\phi_i}^{-1} + \frac{1}{\sigma_i^2}\sum_{t=2}^{T}(h_{i,t-1}-\mu_i)^2\\
	\hat{\phi}_h & = K_{\phi_i}^{-1}\left[V_{\phi_i}^{-1}\phi_{0,i} + \frac{1}{\sigma_i^2}\sum_{t=2}^{T}(h_{i,t-1}-\mu_i) (h_{i,t}-\mu_i) \right].
\end{align*}
Then, given the current draw $\phi_i$, a proposal $\phi_i^*$ is accepted with probability $\min(1,g(\phi_i^*)/g(\phi_i))$; 
otherwise the Markov chain stays at the current state $\phi_i$.

\newpage

\section*{Appendix C: Technical Details on Integrated Likelihood Evaluation}

In this appendix we provide the technical details for evaluating the integrated likelihood outlined in Section \ref{ss:intlike}. Recall that the integrated likelihood can be written as
\begin{equation}\label{eq:intlike2}
	p(\by \gvn\vbeta, \bL,\vmu,\vphi,\vsigma^2)  = \int  p(\by\gvn\vbeta,\bL,\bh) p(\bh \gvn \vmu,\vphi,\vsigma^2) \di\bh,
\end{equation}
where the first term in the integrant has the following expression
\[
	p(\by\gvn\vbeta,\bL,\bh) = (2\pi)^{-\frac{Tn}{2}}\prod_{t=1}^T|\bL\vOmega_t\bL'+\vSigma_t|^{- \frac{1}{2}}
	\e^{-\frac{1}{2}(\by_t - (\mathbf{I}_n\otimes\bx_t')\vbeta)'(\bL\vOmega_t\bL'+\vSigma_t)^{-1}
	(\by_t - (\mathbf{I}_n\otimes\bx_t')\vbeta)}.
\]

Next, we derive the joint density of the log-volatilities $p(\bh \gvn \vmu,\vphi,\vsigma^2)$. To that end, stack the state equations \eqref{eq:ht1}-\eqref{eq:ht2} over $t=1,\ldots, T$:
\[
	\bH_{\vphi}\bh = 	\bH_{\vphi} \bm_{\vmu} + \bu^h, \quad \bu^h\sim\distn{N}(\mathbf{0},\bS_{\vsigma^2}),
\]
where $\bS_{\vsigma^2} = \text{diag}(\sigma_1^2/(1-\phi_1^2), \ldots, \sigma_{n+r}^2/(1-\phi_{n+r}^2),\sigma_1^2,\ldots, \sigma_{n+r}^2,\ldots, \sigma_1^2, \ldots, \sigma_{n+r}^2)'$ and  
\[
	\bm_{\vmu} = \mathbf{1}_T\otimes\begin{pmatrix} \vmu \\ \mathbf{0} \end{pmatrix} \quad
	\bH_{\vphi} = \begin{pmatrix} \mathbf{I}_{n+r} & \mathbf{0} & \cdots & \mathbf{0} \\
				 	 - \diag(\vphi)  & \mathbf{I}_{n+r} & \ddots & \vdots \\
							 \vdots & \ddots & \ddots & \vdots \\
							 \mathbf{0} & \cdots & - \diag(\vphi) & \mathbf{I}_{n+r}
	\end{pmatrix}.
\]
Or equivalently
\[
	\bh = 	\bm_{\vmu} + \bH_{\vphi}^{-1}\bu^h, \quad 
	\bH_{\vphi}^{-1}\bu^h\sim\distn{N}(\mathbf{0},(\bH_{\vphi}'\bS_{\vsigma^2}^{-1}\bH_{\vphi})^{-1}),
\]
as the determinant of the square matrix $\bH_{\vphi}$ is one and is thus invertible. It follows that $(\bh \gvn \vmu,\vphi,\vsigma^2)\sim\distn{N}(\bm_{\vmu},(\bH_{\vphi}'\bS_{\vsigma^2}^{-1}\bH_{\vphi})^{-1})$ with log-density
\begin{equation*}
\begin{split}
	\log p(\bh \gvn \vmu,\vphi,\vsigma^2) & = -\frac{T(n+r)}{2}\log(2\pi) - \frac{T}{2}\sum_{i=1}^{n+r}\log\sigma_i^2 + \frac{1}{2}\sum_{i=1}^{n+r}\log(1-\phi_i^2) \\
	& \quad - \frac{1}{2}(\bh-\bm_{\vmu})'\bH_{\vphi}'\bS_{\vsigma^2}^{-1}\bH_{\vphi}(\bh-\bm_{\vmu}).
\end{split}
\end{equation*}

Next, we introduce an importance sampling estimator to evaluate the integral in \eqref{eq:intlike2}. The ideal zero-variance importance sampling density in this case is the conditional density of $\bh$ given the data and other parameters but marginal of $\bff$, i.e., $p(\bh \gvn \by, \vbeta, \bL, \vmu, \vphi, \vsigma^2)$. But this density cannot be directly used as an importance sampling density as it is non-standard. We instead approximate it using a Gaussian density, which is then used as the importance sampling density.

\subsection*{An EM Algorithm to Obtain the Mode of $p(\bh \gvn \by, \vbeta, \bL, \vmu, \vphi, \vsigma^2)$}

We first develop an EM algorithm to find the maximizer of the log marginal density
$\log p(\bh \gvn \by, \vbeta, \bL, \vmu, \vphi, \vsigma^2)$. To implement the E-step, we compute the following conditional expectation for an arbitrary vector $\breve{\bh}\in\mathbb{R}^{T(n+r)}$:
\[
	\mathcal{Q}(\bh \gvn \breve{\bh}) = \Em_{\bff|\breve{\bh}}\left[ \log p(\bh, \bff \gvn \by, \vbeta, \bL, \vmu, \vphi, \vsigma^2)\right],
\]
where the expectation is taken with respect to $p(\bff \gvn \by, \vbeta, \bL, \breve{\bh}, \vmu, \vphi, \vsigma^2) = p(\bff \gvn \by, \vbeta, \bL, \breve{\bh})$. As discussed in Section~\ref{s:estimation} of  the main text, the latent factors $\bff_1,\ldots, \bff_T$ are conditionally independent given the data and model parameters. In fact, for $t=1,\ldots, T$, they have the following Gaussian distributions: 
\[
	(\bff_t \gvn\by, \vbeta, \bL, \breve{\bh}) \sim  \distn{N}(\hat{\bff_t},\bK_{\bff_t}^{-1}), 
\]
where 
\[
	\bK_{\bff_t} = \breve{\vOmega}^{-1}_t + \bL'\breve{\vSigma}^{-1}_t\bL, \quad	
	\hat{\bff_t}  = \bK_{\bff_t}^{-1}\bL'\breve{\vSigma}^{-1}_t(\by_t- (\mathbf{I}_n\otimes \bx_t')\vbeta).
\]
Note that here we use  $\breve{\bh} = (\breve{\bh}_{1}^{y'},\breve{\bh}_{1}^{f'}, \ldots,\breve{\bh}_{T}^{y'},\breve{\bh}_{T}^{f'})' = (\breve{h}_{1,1},\ldots,\breve{h}_{n+r,1},\ldots, \breve{h}_{1,T},\ldots,\breve{h}_{n+r,T})'$ to construct $\breve{\vSigma}_t = \text{diag}(\breve{\bh}_{t}^{y}) = \text{diag}(\e^{\breve{h}_{1,t}},\ldots, \e^{\breve{h}_{n,t}})$ and $\breve{\vOmega}_t = \text{diag}(\breve{\bh}_{t}^{f}) = \text{diag}(\e^{\breve{h}_{n+1,t}},\ldots, \e^{\breve{h}_{n+r,t}})$ instead of $\bh$.
 
Then, an explicit expression of $\mathcal{Q}(\bh \gvn \breve{\bh})$ can be derived as follows:
\begin{align*}
	\mathcal{Q}(\bh\gvn \breve{\bh}) = & -\frac{1}{2}(\bh-\bm_{\vmu})'\bH_{\vphi}'\bS_{\vsigma^2}^{-1}\bH_{\vphi}(\bh-\bm_{\vmu}) -\frac{1}{2}\mathbf{1}_{T(n+r)}'\bh \\	
	& -\frac{1}{2}\sum_{t=1}^T \Em_{\bff|\breve{\bh}}
		\left[\bff_t'\vOmega^{-1}_t \bff_t + (\vepsilon_t-\bL\bff_t)'\vSigma^{-1}_t
			(\vepsilon_t-\bL\bff_t)\right] + c_1  \\
= & -\frac{1}{2}(\bh-\bm_{\vmu})'\bH_{\vphi}'\bS_{\vsigma^2}^{-1}\bH_{\vphi}(\bh-\bm_{\vmu}) 
-\frac{1}{2} \mathbf{1}_{T(n+r)}'\bh -\frac{1}{2}\sum_{t=1}^T\text{tr}\left(\diag(\e^{-\bh_t^f})(\hat{\bff}_t\hat{\bff}_t'+\bK_{\bff_t}^{-1})\right)  \\	
	& -\frac{1}{2}\sum_{t=1}^T\text{tr}\left(\diag(\e^{-\bh_t^y})\left((\vepsilon_t-\bL\hat{\bff}_t)(\vepsilon_t-\bL\hat{\bff}_t)' + \bL\bK_{\bff_t}^{-1}\bL'\right)\right) + c_1,
\end{align*}
where $\vepsilon_t = \by_t- (\mathbf{I}_n\otimes \bx_t')\vbeta$ and $c_1$ is a constant not dependent on $\bh$.

In the M-step, we maximize the function $\mathcal{Q}(\bh \gvn \breve{\bh}) $ with respect to $\bh$. This can be done using the Newton-Raphson method \citep[see, e.g.,][]{handbook11}.
To compute the gradient and Hessian of $\mathcal{Q}(\bh \gvn \breve{\bh})$, let $\hat{z}_{i,t}^y$ denote the $i$-th diagonal element of $(\vepsilon_t-\bL\hat{\bff}_t)(\vepsilon_t-\bL\hat{\bff}_t)' + \bL\bK_{\bff_t}^{-1}\bL'$, $i=1,\ldots,n$. Similarly, let $\hat{z}_{j,t}^f$ denote the $j$-th diagonal element of $(\hat{\bff}_t\hat{\bff}_t'+\bK_{\bff_t}^{-1})$, $j=1\ldots, r$. Finally, define $\hat{\bz} = (\hat{\bz}_1',\ldots, \hat{\bz}_T')'$, where $\hat{\bz}_t = (\hat{z}_{1,t}^y,\ldots, \hat{z}_{n,t}^y, \hat{z}_{1,t}^f, \ldots, \hat{z}_{r,t}^f)'$. Then, we can rewrite  $\mathcal{Q}(\bh \gvn \breve{\bh}) $  more compactly as
\[
	\mathcal{Q}(\bh\gvn \breve{\bh}) = -\frac{1}{2}(\bh-\bm_{\vmu})'\bH_{\vphi}'\bS_{\vsigma^2}^{-1}\bH_{\vphi}(\bh-\bm_{\vmu}) -\frac{1}{2} \mathbf{1}_{T(n+r)}'\bh 
	-\frac{1}{2}\hat{\bz}' \e^{-\bh}.
\]
Hence, the gradient is given by
\[
	\bg_\mathcal{Q} = -\bH_{\vphi}'\bS_{\vsigma^2}^{-1}\bH_{\vphi}(\bh-\bm_{\vmu})
	-\frac{1}{2}\left(\mathbf{1}_{T(n+r)} -\e^{-\bh}\odot \hat{\bz}\right),
\]
and the Hessian is
\begin{equation}\label{eq:Hess_Q}
	\bH_\mathcal{Q} = - \bH_{\vphi}'\bS_{\vsigma^2}^{-1}\bH_{\vphi} 
	- \frac{1}{2}\diag\left(\e^{-\bh}\odot \hat{\bz}\right),
\end{equation}
where $\odot$ denotes the entry-wise product. Since the determinant $|\bH_{\vphi}'\bS_{\vsigma^2}^{-1}\bH_{\vphi}| = |\bS_{\vsigma^2}|^{-1}$ is strictly positive and the diagonal elements of $\diag\left(\e^{-\bh}\odot \hat{\bz}\right)$ are positive, the Hessian $\bH_\mathcal{Q}$ is negative definite for all $\bh\in\mathbf{R}^{T(n+r)}$. This guarantees fast convergence of the Newton-Raphson method. In addition, the Hessian is a band matrix. This property can be used to further speed up computations with sparse and band matrix routines.

Given the E- and M-steps above, the EM algorithm can be implemented as follows.
We initialize the algorithm with $\bh = \bh^{(0)}$ for some constant vector $\bh^{(0)}$. At the $j$-th iteration, we obtain $\bg_\mathcal{Q}$ and $\bH_\mathcal{Q}$, where 
$\hat{\bff}_t$ and $\bK_{\bff_t}, t=1,\ldots, T,$ are evaluated using $ \bh^{(j-1)}$.
 Then, we compute
\[
	\bh^{(j)} = \argmax_{\bh} \mathcal{Q}(\bh\gvn \bh^{(j-1)})
\]
using the Newton-Raphson method. We repeat the E- and M-steps until some convergence criterion is met, e.g.,
the norm between consecutive $\bh^{(j)}$ is less than a pre-determined tolerance value.
At the end of the EM algorithm, we obtain the mode of the density $p(\bh \gvn \by, \vbeta, \bL, \vmu, \vphi, \vsigma^2)$, which is denoted by $\hat{\bh}$. We summarize the EM algorithm in Algorithm~\ref{alg:EM}.

\begin{algorithm}[H]
\caption{EM algorithm to obtain the mode of $p(\bh \gvn \by, \vbeta, \bL, \vmu, \vphi, \vsigma^2)$.}
\label{alg:EM}
Suppose we have an initial guess $\bh^{(0)}$ and error tolerance levels $\epsilon_1$ and $\epsilon_2$, say, $\epsilon_1 = \epsilon_2 = 10^{-4}$.
The EM algorithm consists of iterating the following steps for
$j=1,2,\ldots$:
\begin{enumerate}
	\item E-Step: Given the current value $\bh^{(j-1)}$, compute $\mathbf{K}_{\bff_t}$, $\hat{\bff_t}, t=1,\ldots, T,$ and $\hat{\bz}$
	
	\item M-Step: Maximize $\mathcal{Q}(\bh \gvn \bh^{(j-1)})$ with respect to $\bh$ by the Newton-Raphson method.
    That is, set $\bh^{(0,j-1)} = \bh^{(j-1)}$ and iterate the following steps for  $k=1,2,\ldots$:
	
		\begin{enumerate}			
			\item Compute $\bg_\mathcal{Q}$ and $\bH_\mathcal{Q}$ using
					$\mathbf{K}_{\bff_t}$, $\hat{\bff_t}, t=1,\ldots, T,$ and $\hat{\bz}$ 
					obtained in the E-step, and set $\bh = \bh^{(k-1,j-1)}$
			
			\item Update $\bh^{(k,j-1)} = \bh^{(k-1,j-1)} - \bH_\mathcal{Q}^{-1}\bg_\mathcal{Q}$
			
			\item If, for example, $ \|\bh^{(k,j-1)}-\bh^{(k-1,j-1)}\| < \epsilon_1$, terminate the iteration and set
			 $\bh^{(j)} = \bh^{(k,j-1)}$.
		\end{enumerate}
	
	\item Stopping condition: if, for example, $\|\bh^{(j)}-\bh^{(j-1)}\| < \epsilon_2$, terminate the algorithm.
\end{enumerate}
\end{algorithm}

\subsection*{Computing the Hessian of $\log p(\bh \gvn \by, \vbeta, \bL, \vmu, \vphi, \vsigma^2)$}

After obtaining the mode $\hat{\bh}$ of the log density $\log p(\bh \gvn \by, \vbeta, \bL, \vmu, \vphi, \vsigma^2)$, next we compute the Hessian evaluated at $\hat{\bh}$. Here we describe two approaches to do so. In the first approach, we provide an approximation of the Hessian using the EM algorithm. The resulting matrix is banded and is guaranteed to be negative definite. In the second approach, we directly compute the Hessian of $\log p(\bh \gvn \by, \vbeta, \bL, \vmu, \vphi, \vsigma^2)$. In our experience the two approaches give very similar results, but the first approach is more numerically stable.

In what follows, we start with the first approach. Note that by Bayes' theorem, we have
\[
	p(\bh \gvn \by, \vbeta, \bL, \vmu, \vphi, \vsigma^2) = 
	\frac{p(\bh,\bff \gvn \by, \vbeta, \bL, \vmu, \vphi, \vsigma^2)}
	{p(\bff \gvn \bh, \by, \vbeta, \bL, \vmu, \vphi, \vsigma^2)}.
\]
If we take the log of both sides and then take expectation with respect to
$p(\bff \gvn \bh, \by, \vbeta, \bL)$, we obtain the identity
\begin{equation}\label{eq:h_QH}
	\log p(\bh \gvn \by, \vbeta, \bL, \vmu, \vphi, \vsigma^2) = \mathcal{Q}(\bh \gvn \bh) + \mathcal{H}(\bh \gvn \bh),
\end{equation}
where $\mathcal{H}(\bh \gvn \bh) = -\Em_{\bff| \bh}\left[ \log p(\bff \gvn \bh, \by, \vbeta, \bL, \vmu, \vphi, \vsigma^2)\right] = -\Em_{\bff| \bh}\left[ \log p(\bff \gvn \bh, \by, \vbeta, \bL, \vmu)\right]$.

It follows that the Hessian of $\log p(\bh \gvn \by, \vbeta, \bL, \vmu, \vphi, \vsigma^2)$ evaluated at $\hat{\bh}$ is simply the sum of the Hessians of $\mathcal{Q}$ and $\mathcal{H}$ with $\bh = \hat{\bh}$. Note that the Hessian of  $\mathcal{Q}(\bh \gvn \hat{\bh})$ comes out as a by-product of the EM algorithm; an analytical expression is given in \eqref{eq:Hess_Q}. We use it as an approximation of the Hessian of  $\mathcal{Q}(\bh \gvn \bh)$ evaluated at $\bh = \hat{\bh}$. Next, we derive an analytical expression for $\mathcal{H}(\bh \gvn \bh)$:
\begin{align*}
	\mathcal{H}(\bh \gvn \bh) &
	= -\Em_{\bff| \bh}\left[ \log p(\bff \gvn \bh, \by, \vbeta, \bL, \vmu) \right] \\
	& = \frac{Tr}{2}\log(2\pi)  -\frac{1}{2}\sum_{t=1}^T \log|\bK_{\bff_t}| + \frac{1}{2}
	\sum_{t=1}^T \Em_{\bff_t|\bh}\left[(\bff_t-\hat{\bff}_t)'\bK_{\bff_t}
	(\bff_t-\hat{\bff}_t)\right] \\
	& = -\frac{1}{2}\sum_{t=1}^T\log|\bL'\text{diag}(\e^{-\bh_t^y})\bL + \text{diag}(\e^{-\bh_t^f})|+ c_2 \\
	& = -\frac{1}{2}\sum_{t=1}^T\log|\bW'\text{diag}(\e^{-\bh_t})\bW| + c_2,
\end{align*}
where $c_2$ is a constant not dependent on $\bh$ and $\bW = \begin{pmatrix} \bL \\ \mathbf{I}_r \end{pmatrix}$. In the above derivation we have used the fact that under 
$p(\bff_t \gvn \bh, \by, \vbeta, \bL, \vmu)$, the quadratic form
$(\bff_t-\hat{\bff}_t)'\bK_{\bff_t}(\bff_t-\hat{\bff}_t)$ is a chi-squared random variable and its expectation does not depend on $\bh$ (and thus absorbed into the constant $c_2$).

To compute the Hessian of $\mathcal{H}$, we first note that
\begin{align*}
	\frac{\partial}{\partial h_{i,t}}\bK_{\bff_t} & = \frac{\partial}{\partial h_{i,t}} 
	\bW'\text{diag}(\e^{-\bh_t})\bW = \frac{\partial}{\partial h_{i,t}} \sum_{j=1}^{n+r}
	\e^{-h_{j,t}}\bw_j\bw_j' = -\e^{-h_{i,t}}\bw_i\bw_i', \\
	\frac{\partial}{\partial h_{i,s}}\bK_{\bff_t} & = 0, \quad s\neq t,
\end{align*}
where $\bw_j'$ is the $j$-th row of $\bW$. Next, using standard results of matrix differentiation, we obtain
\begin{align*}
	\frac{\partial}{\partial h_{i,t}} \mathcal{H}(\bh \gvn \bh) & = -\frac{1}{2}\text{tr}\left(\bK_{\bff_t}^{-1} \frac{\partial\bK_{\bff_t}}{\partial h_{i,t}}\right) 
	= \frac{1}{2}\e^{-h_{i,t}}\bw_i'\bK_{\bff_t}^{-1}\bw_i, \\
	\frac{\partial^2}{\partial h_{i,t}^2} \mathcal{H}(\bh \gvn \bh) & = -\frac{1}{2}\left(\e^{-h_{i,t}}\bw_i'\bK_{\bff_t}^{-1}\bw_i + \e^{-h_{i,t}}\bw_i' \bK_{\bff_t}^{-1}\frac{\partial \bK_{\bff_t}}{\partial h_{i,t}} \bK_{\bff_t}^{-1}\bw_i\right)  \\	
	& = -\frac{1}{2}\e^{-h_{i,t}}\bw_i'
	\bK_{\bff_t}^{-1}\bw_i( 1 - \e^{-h_{i,t}}\bw_i'\bK_{\bff_t}^{-1}\bw_i),  \\
	\frac{\partial^2}{\partial h_{i,t}\partial h_{j,t}} \mathcal{H}(\bh \gvn \bh) & = \frac{1}{2} \e^{-(h_{i,t}+h_{j,t})}\bw_i'	\bK_{\bff_t}^{-1}\bw_j\bw_j'	\bK_{\bff_t}^{-1}\bw_i, \quad i\neq j, \\
	\frac{\partial^2}{\partial h_{i,t}\partial h_{j,s}} \mathcal{H}(\bh \gvn \bh) & = 0, 
	\quad s\neq t.	
\end{align*}
Hence, the Hessian is block diagonal (and hence banded). More specifically, the Hessian of $\mathcal{H}(\bh \gvn \bh)$ can be written in the following matrix form
\[
	\bH_\mathcal{H} = -\frac{1}{2}\bZ' \odot (\mathbf{I}_{T(n+r)} - \bZ),
\]
where $\bZ = \text{diag}(\bZ_1,\ldots, \bZ_T)$ with 
$\bZ_t = \text{diag}(\e^{-\bh_t})\bW\bK_{\bff_t}^{-1} \bW'.$

Finally, let $\bH_\mathcal{Q}$ denote the  Hessian of $\mathcal{Q}(\bh\gvn\bh)$ evaluated at $\bh = \hat{\bh}$. Then, the negative Hessian of the log marginal density of $\bh$ evaluated at $\bh = \hat{\bh}$ is simply
$\bK_{\bh} = - (\bH_\mathcal{Q} + \bH_\mathcal{H})$, which is used as the precision matrix of the Gaussian approximation. Note that since both $\bH_\mathcal{Q}$ and $\bH_\mathcal{H}$ are band matrices, so is $\bK_{\bh}$.

The second approach directly computes the Hessian of the log marginal density:
\begin{align*}
   \log p(\bh \gvn \by,  \vbeta, \bL, \vmu, \vphi,\vsigma^2) & = c_3 + \log p(\by \gvn \vbeta, \bh, \bL) 
	 + \log p(\bh \gvn \vmu, \vphi,\vsigma^2),  \\
	 & = c_4 \underbrace{-\frac{1}{2}\sum_{t=1}^T \log |\bL\vOmega_t\bL'+\vSigma_t|}_{T_1(\bh)} \underbrace{-\frac{1}{2}\sum_{t=1}^T \vepsilon_t'(\bL\vOmega_t\bL'+\vSigma_t)^{-1}\vepsilon_t}_{T_2(\bh)} \\
			& \qquad \underbrace{-\frac{1}{2}(\bh-\bm_{\vmu})'\bH_{\vphi}'\bS_{\vsigma^2}^{-1}\bH_{\vphi}(\bh-\bm_{\vmu})}_{T_3(\bh)},
\end{align*}
where  $\vepsilon_t=\by_t - (\mathbf{I}_n\otimes\bx_t')\vbeta$, and $c_3$ and  $c_4$ are constants not dependent on $\bh$. Next we derive the Hessians of the functions $T_1, T_2$ and $T_3$.

Let $\tilde{\bW} = \begin{pmatrix} \bL' \\ \mathbf{I}_n \end{pmatrix}$. Then, $\bG_t \equiv \bL\vOmega_t\bL'+\vSigma_t = \tilde{\bW}'\text{diag}(\e^{\bh_t})\tilde{\bW}$. Using a similar derivation of $\bH_\mathcal{H}$ in the EM algorithm, it is easy to see that the Hessian of $ T_1(\bh)$ is given by:
\[
  \bH_{T_1(\bh)} = -\frac{1}{2}\tilde{\bZ}' \odot (\mathbf{I}_{T(n+r)} - \tilde{\bZ}),
\]
where $\tilde{\bZ} = \text{diag}(\tilde{\bZ}_1,\ldots, \tilde{\bZ}_T)$ with 
$\tilde{\bZ}_t = \text{diag}(\e^{\bh_t})\tilde{\bW}\bG_{t}^{-1} \tilde{\bW}'$. It is also clear that 
the Hessian of $T_3(\bh)$ is simply
\[
   \bH_{T_3(\bh)} =-\bH_{\vphi}'\bS_{\vsigma^2}^{-1}\bH_{\vphi}.
\]
Next, we derive the Hessian of $T_2(\bh)$ below. First note that
\begin{align*}
	\frac{\partial}{\partial h_{i,t}}\bG_t & = \frac{\partial}{\partial h_{i,t}} 
	\tilde{\bW}'\text{diag}(\e^{\bh_t})\tilde{\bW} = \frac{\partial}{\partial h_{i,t}} \sum_{j=1}^{n+r}
	\e^{h_{j,t}}\tilde{\bw}_j\tilde{\bw}_j' = \e^{h_{i,t}}\tilde{\bw}_i\tilde{\bw}_i', \\
	\frac{\partial}{\partial h_{i,s}}\bG_t & = 0, \quad s\neq t,
\end{align*}
where $\tilde{\bw}_j'$ is the $j$-th row of $\tilde{\bW}$. Next, using standard results of matrix differentiation, we obtain
\begin{align*}
	\frac{\partial}{\partial h_{i,t}} T_2(\bh) & = \frac{1}{2} \vepsilon_t' \bG_t^{-1}\frac{\partial \bG_t}{\partial h_{i,t}} \bG_t^{-1}\vepsilon_t=\frac{1}{2} \e^{h_{i,t}}\vepsilon_t' \bG_t^{-1}\tilde{\bw}_i\tilde{\bw}_i' \bG_t^{-1}\vepsilon_t=\frac{1}{2} \e^{h_{i,t}}(\vepsilon_t' \bG_t^{-1}\tilde{\bw}_i)^2, \\
	\frac{\partial^2}{\partial h_{i,t}^2} T_2(\bh) & = \frac{1}{2}\left(\e^{h_{i,t}}(\vepsilon_t' \bG_t^{-1}\tilde{\bw}_i)^2 - 2\e^{h_{i,t}}(\vepsilon_t' \bG_t^{-1}\tilde{\bw}_i)\vepsilon_t' \bG_t^{-1}\frac{\partial \bG_t}{\partial h_{i,t}} \bG_t^{-1}\tilde{\bw}_i \right)  \\	
	& = \frac{1}{2}\e^{h_{i,t}}(\vepsilon_t' \bG_t^{-1}\tilde{\bw}_i)^2\left( 1 - 2\e^{h_{i,t}} \tilde{\bw}_i' \bG_t^{-1}\tilde{\bw}_i\right),  \\
	\frac{\partial^2}{\partial h_{i,t}\partial h_{j,t}} T_2(\bh) & = -\e^{(h_{i,t}+h_{j,t})}(\vepsilon_t' \bG_t^{-1}\tilde{\bw}_i)(\vepsilon_t' \bG_t^{-1}\tilde{\bw}_j)(\tilde{\bw}_j' \bG_t^{-1}\tilde{\bw}_i), \quad i\neq j, \\
	\frac{\partial^2}{\partial h_{i,t}\partial h_{j,s}} T_2(\bh) & = 0, 
	\quad s\neq t.	
\end{align*}
More specifically, the Hessian of $T_2(\bh)$ can be written in the following matrix form
\[
	\bH_{T_2(\bh)} = \frac{1}{2}\breve{\bZ}' \odot (\mathbf{I}_{T(n+r)} - 2\tilde{\bZ}),
\]
where $\breve{\bZ} = \text{diag}(\breve{\bZ}_1,\ldots, \breve{\bZ}_T)$ with 
$\breve{\bZ}_t = \text{diag}(\e^{\bh_t})\tilde{\bW}\bG_{t}^{-1}\vepsilon_t\vepsilon_t'\bG_{t}^{-1} \tilde{\bW}'$.
Finally, the Hessian from direct computation is simply $\bH_{\rm Direct}= \bH_{T_1(\bh)}+ \bH_{T_2(\bh)} + \bH_{T_3(\bh)}$. Since $\bH_{T_1(\bh)}, \bH_{T_2(\bh)}$ and $\bH_{T_3(\bh)}$ are all band matrices, so is $\bH_{\rm Direct}$.

\newpage

\section*{Appendix D: Additional Monte Carlo Results} \label{app:D}

In this appendix we present results on two artificial data experiments to illustrate the estimation accuracy of the VAR-FSV model under DGPs with and without stochastic volatility. In the first experiment, we generate a dataset from the VAR-FSV in \eqref{eq:yt}--\eqref{eq:ht2} with $n=10$, $T=500$, $r=3$ factors and $p=4$ lags. We then estimate the model using the posterior sampler outlined in Section~\ref{s:estimation}. The first dataset is generated as follows. First, the intercepts are drawn independently from the uniform distribution on the interval $(-10,10)$, i.e., $\distn{U}(-10, 10)$. For the VAR coefficients, the diagonal elements of the first VAR coefficient matrix are iid $\distn{U}(0,0.5)$ and the off-diagonal elements are from $\distn{U}(-0.2,0.2)$; all other elements of the $j$-th ($j > 1$) VAR coefficient matrices are iid $\distn{N}(0,0.1^2/j^2).$ All elements of the factor loadings matrix are iid standard normal: $\bL_{ij} \sim \mathcal{N}(0, 1)$ for $i=1,\ldots,n$ and $j=1,\ldots,r$. Finally, for the log-volatility processes, we set $\mu_i=-1, \phi_i=0.98$ and $\sigma_i=0.1$ for $i=1,\ldots,n$, and $\phi_{n+j} = 0.98$, $\sigma_{n+j}=0.1$ for $j=1,\ldots,r$. 

\begin{figure}[H]
	\begin{center} 
		\includegraphics[width=.95\textwidth]{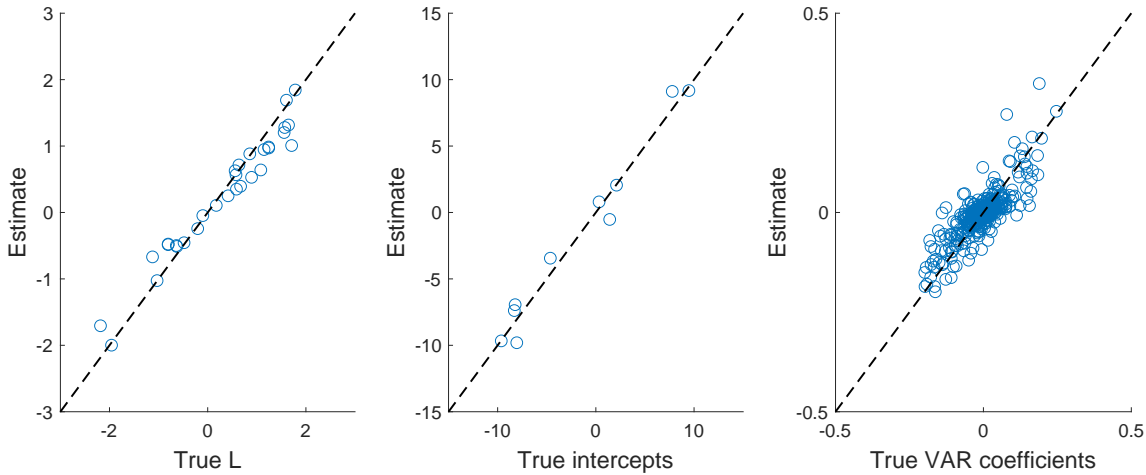} 
	\end{center}
	\caption{Scatter plots of the posterior means of the factor loadings (left panel), intercepts (middle panel) and VAR coefficients (right panel) against the true values from a DGP with stochastic volatility.}
\label{fig:sim1_coef}
\end{figure}

The results of the artificial data experiments are reported in Figures~\ref{fig:sim1_coef}-\ref{fig:sim1_h}. It is evident from the figures that the posterior sampler works well and the posterior means track the true values closely.

\begin{figure}[H]
	\begin{center}
		\includegraphics[width=.6\textwidth]{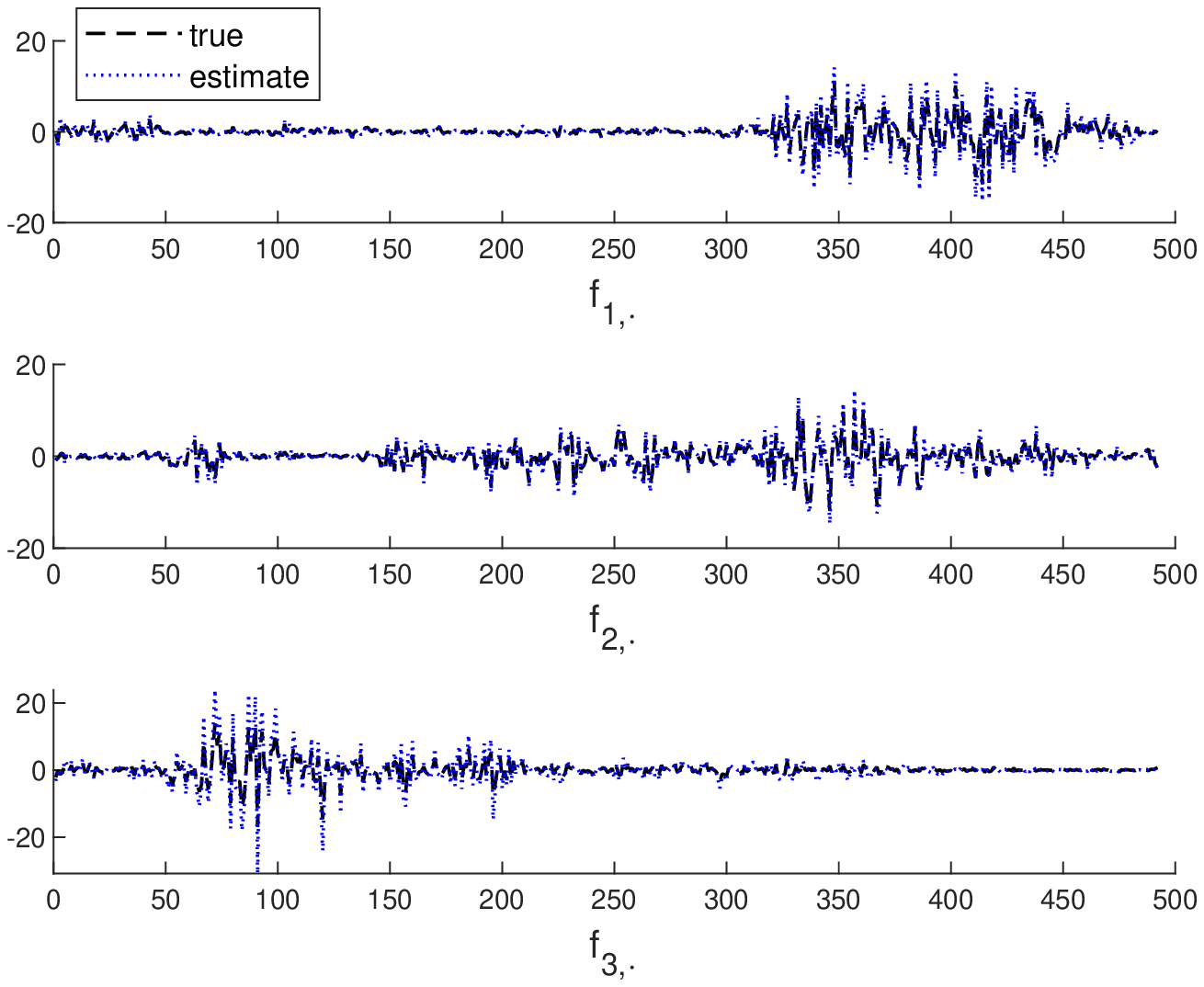} 
	\end{center}
	\caption{Time series plots of the posterior means of the factors $\bff_{i,\cdot}$, $i=1,2,3,$ from a DGP with stochastic volatility.}
\end{figure}

\begin{figure}[H] 
	\begin{center}
		\includegraphics[width=.6\textwidth]{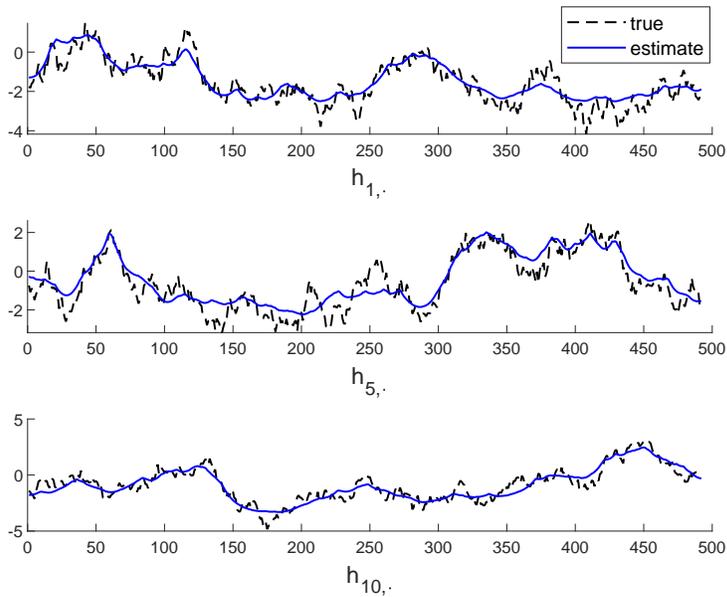} 
	\end{center}
	\caption{Time series plots of the posterior means of the stochastic volatilities $\bh_{i,\cdot}$, $i=1,5,10,$ from a DGP with stochastic volatility.}
	\label{fig:sim1_h}
\end{figure}

\begin{figure}[H]
	\begin{center}
		\includegraphics[width=.95\textwidth]{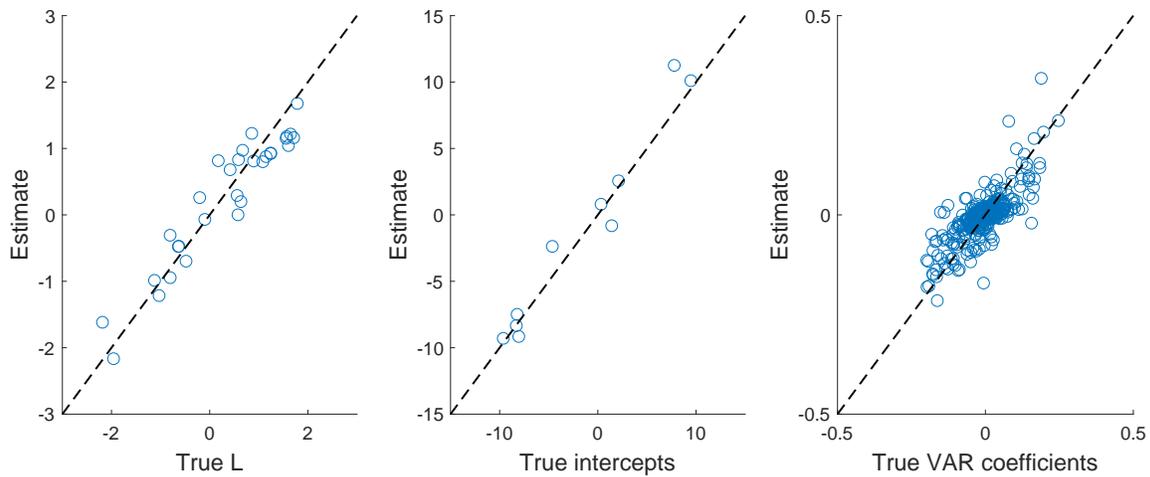} 
	\end{center}
	\caption{Scatter plots of the posterior means of the factor loadings (left panel), intercepts (middle panel) and VAR coefficients (right panel) against the true values from a DGP with partial stochastic volatility.}
\label{fig:sim2_coef}

\end{figure}

In the second experiment, we generate data from the same VAR-FSV but with several stochastic volatility components turned off. In particular, we set $h_{i,t} = 0$ for $i=1,2,3,4,5$, and $h_{n+j,t} = 0$ for $j=2,3$. That is, the idiosyncratic errors of the first five variables, as well as the last two factors, are homoscedastic. We then fit the data using the (mis-specified) fully heteroscedastic model. The results are reported in Figures~\ref{fig:sim2_coef}-\ref{fig:sim2_h}. When the DGP does not have full stochastic volatility, some elements of the factor loading matrix $\bL$ are harder to pin down, since they are not point-identified. But it is interesting to note that the estimates of the stochastic volatility are still able to track the true values fairly closely. The estimates of the VAR coefficients are also close to the true values. 

\begin{figure}[H]
	\begin{center}
		\includegraphics[width=.6\textwidth]{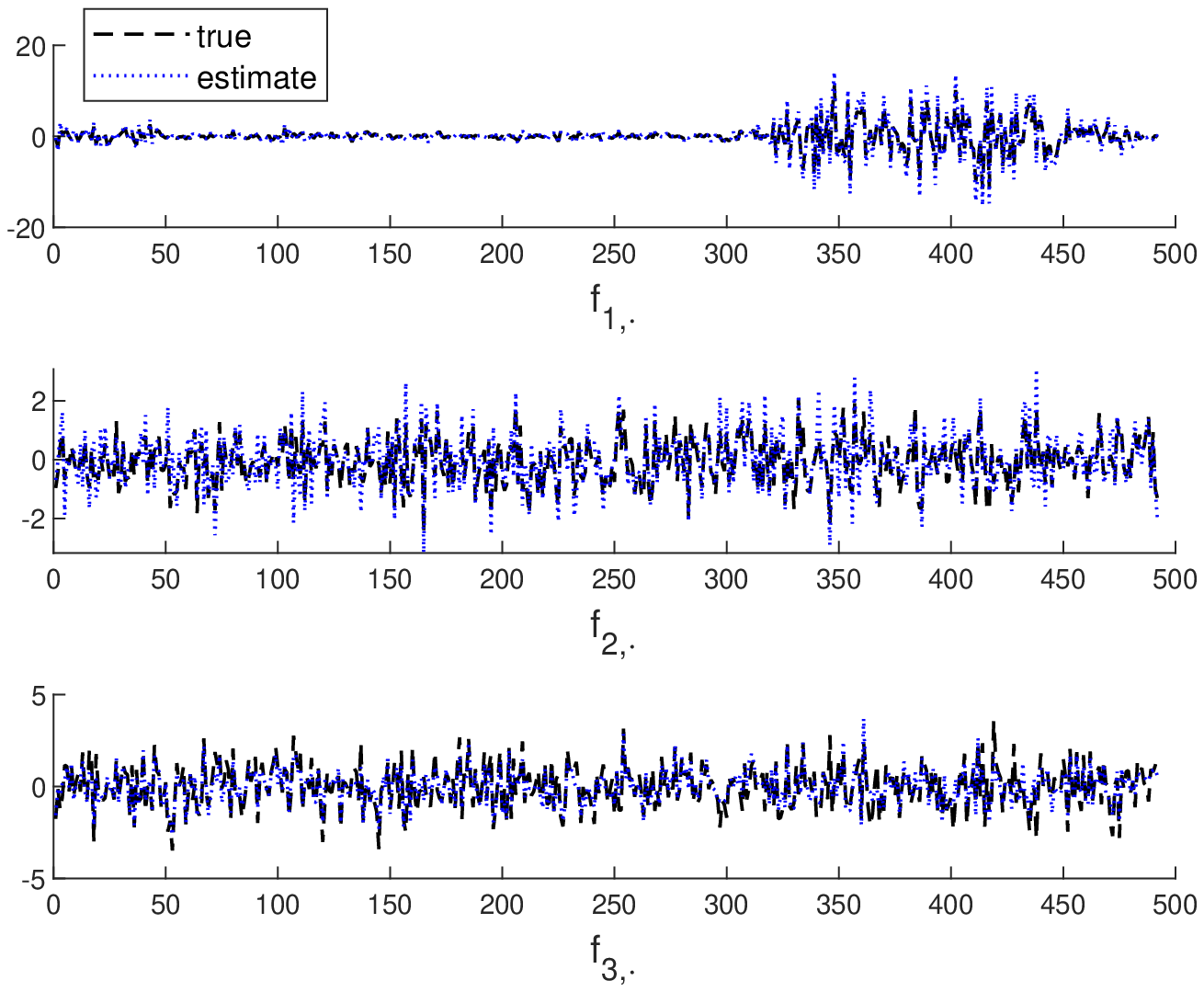} 
	\end{center}
	\caption{Time series plots of the posterior means of the factors $\bff_{i,\cdot}$, $i=1,2,3,$ from a DGP with partial stochastic volatility.}
\end{figure}

\begin{figure}[H]
	\begin{center}
		\includegraphics[width=.6\textwidth]{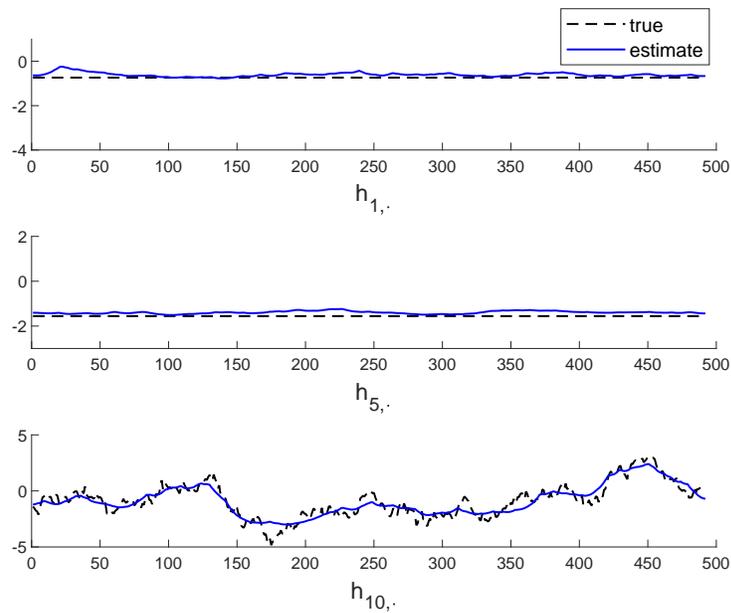} 
	\end{center}
		\caption{Time series plots of the posterior means of the stochastic volatilities $\bh_{i,\cdot}$, $i=1,5,10,$ from a DGP with partial stochastic volatility.}
	\label{fig:sim2_h}
\end{figure}

\newpage

\section*{Appendix E: Data}

This appendix provides the details of the raw data used to construct the variables in the empirical application. In particular, Table~\ref{tab:var} lists the variables and their sources. The sample period is from 1985:Q1 to 2013:Q2.
\begin{table}[H]
\centering
\caption{Description of variables used in the empirical application.} \label{tab:var}
\resizebox{\textwidth}{!}{\begin{tabular}{lll}
\hline\hline
Variable 	&	Description	&	Source	\\ \hline
GDP	&	Log of real GNP/GDP	&	Federal Reserve Bank of Philadelphia	\\ 
\rowcolor{lightgray}
GDP Deflator	&	Log of price index of GNP/GDP 	&	Federal Reserve Bank of Philadelphia	\\
3-month treasury bill	&	3-month treasury bill rate	&	Federal Reserve Bank of St. Louis	\\
\rowcolor{lightgray}
Investment	&	Log of real gross private domestic investment	&	Federal Reserve Bank of St. Louis	\\
S\&P 500	&	Log of S\&P 500	&	Yahoo Finance	\\
\rowcolor{lightgray}
Total credit	&	Log of loans to non-financial private sector	&	Board of Governors of the Federal \\
\rowcolor{lightgray}
& & Reserve System	\\
Mortgages	&	Log of home mortgages of households and & Federal Reserve Bank of St. Louis \\
  & non-profit organizations	&		\\
\rowcolor{lightgray}
Real personal consumption expenditures	&	Log of real personal consumption expenditures & Federal Reserve Bank of St. Louis \\ 
Real estate value	&	Log of real estate at market value of households & Federal Reserve Bank of St. Louis \\ 
 & and non-profit organizations	&		\\
\rowcolor{lightgray}
Corporate bond yield	&	Moody's baa corporate bond yield	&	Federal Reserve Bank of St. Louis	\\
10-year treasury note	&	10-year treasury constant maturity rate	&	Federal Reserve Bank of St. Louis	\\
\rowcolor{lightgray}
Federal funds rate	&	Federal funds rate	&	Federal Reserve Bank of St. Louis	\\
Mortgage rate	&	30-year fixed rate mortgage average	&	Federal Reserve Bank of St. Louis	\\
\rowcolor{lightgray}
CPI	&	Log of consumer price index	&	Federal Reserve Bank of St. Louis	\\
PCE	&	Log of price index of personal consumption &	Federal Reserve Bank of St. Louis	\\
& expenditure & \\
\rowcolor{lightgray}
Employment	&	Log of employment level	&	Federal Reserve Bank of St. Louis	\\
All employees: manufacturing	&	Log of all employees in the manufacturing sector	&	Federal Reserve Bank of St. Louis	\\
\rowcolor{lightgray}
Industrial production	&	Log of industrial production index	&	Federal Reserve Bank of St. Louis	\\
Industrial production: final products	&	Log of industrial production: final products index	&	Federal Reserve Bank of St. Louis	\\
\rowcolor{lightgray}
1-year treasury bill	&	1-year treasury constant maturity rate	&	Federal Reserve Bank of St. Louis	\\
Dow Jones Industrial Average	&	Log of Dow Jones Industrial Average index	&	Google Finance	\\
\rowcolor{lightgray}
Nasdaq Composite	&	Log of Nasdaq Composite	&	Federal Reserve Bank of St. Louis	\\
\hline\hline
\end{tabular}
}
\end{table}

\newpage

\section*{Appendix F: Structural Analysis Tools}

In this appendix we provide details on various structural analysis tools for the VAR-FSV similar to those designed for the structural VAR. In what follows, we describe methods to construct structural impulse response functions, forecast error variance decompositions and historical decompositions.

To derive expressions of responses of $\by_t$ to a one-time impulse in $\mathbf{f}_t$, we first rewrite the VAR($p$) in \eqref{eq:yt} as an equivalent VAR(1) as follows:
\[
	\bY_t = \bA_0 + \bA \bY_{t-1} +\bE_t, 
\]
where 
\[
\bY_t =\begin{pmatrix}
\by_t  \\
\by_{t-1} \\
\vdots \\
\by_{t-p+1}  \\
\end{pmatrix},  \ \ 
\bA_0 =\begin{pmatrix}
\ba_0 \\
\mathbf{0} \\
\vdots \\
\mathbf{0} \\
\end{pmatrix},  \ \ 
\bE_t =\begin{pmatrix}
\vepsilon_t \\
\mathbf{0} \\
\vdots \\
\mathbf{0}
\end{pmatrix},  \ \ 
\bA  = \begin{pmatrix}
\bA_1 & \bA_2 & \cdots & \bA_{p-1} & \bA_p\\
\mathbf{I}_n  & \mathbf{0} & \cdots & \mathbf{0} & \mathbf{0}\\
\mathbf{0} & \mathbf{I}_n  & \mathbf{0} & \cdots & \mathbf{0}\\
\vdots & \ddots & \ddots & \ddots & \vdots \\
\mathbf{0} & \cdots & \mathbf{0} & \mathbf{I}_n & \mathbf{0}
\end{pmatrix}. 	
\]
By successive substitution for $\bY_{t-s}$, this VAR(1) has a vector moving average representation \citep[see, e.g., Section 4.1 of][]{KL17}:
\begin{equation}
	\bY_t       = \sum_{s=0}^{\infty} \bA^{s} \bA_0+  \sum_{s=0}^{\infty} \bA^{s} \bE_{t-s}= (\mathbf{I}_{np}-\bA)^{-1} \bA_0+  \sum_{s=0}^{\infty} \bA^{s} \bE_{t-s}. \label{eq:yt3} 
\end{equation}
Left-multiplying~\eqref{eq:yt3} by $\bJ\equiv \left(\bfI_n, \mathbf{0}_{n\times n(p-1)}\right)$, which is of dimension $n \times np$, we have
\begin{align*}
	\by_t & = (\bfI_n-\bA_1-\cdots-\bA_p)^{-1} \ba_0+  \sum_{s=0}^{\infty} 
	\bJ\bA^{s}\bJ' \bJ \bE_{t-s} \\
	& = \vmu_{\by}+  \sum_{s=0}^{\infty} \vPhi_s \vepsilon_{t-s}, 
\end{align*}
where $\vmu_{\by} = (\bfI_n-\bA_1-\ldots-\bA_p)^{-1} \ba_0$ and $\vPhi_s=\bJ\bA^{s}\bJ' 
= [\bA^{s}]_{1:n,1:n}$ is the first $n\times n$ block of $\bA^{s}$. Note that the coefficient matrices $\vPhi_s$ can also be calculated recursively as
\[
	\vPhi_0 = \bfI_n, \quad \text{and} \quad \vPhi_s = \sum_{j=0}^{s} \vPhi_{s-j} \bA_j, \quad s=1,2,\ldots
\]
with $\bA_j = \mathbf{0}$ for $j>p$. Finally, using the factor structure in \eqref{eq:epsilont} and standardizing the factors via $\tilde{\mathbf{f}}_{t} = \vOmega_t^{-\frac{1}{2}}\mathbf{f}_t$ with $\vOmega_t = \text{diag}(\e^{h_{n+1,t}},\ldots, \e^{h_{n+r,t}})$ so that $\tilde{\mathbf{f}}_{t} \sim\distn{N}(\mathbf{0},\mathbf{I}_r)$, we thus obtain
\begin{equation}\label{eq:VMA}
	\by_t = \vmu_{\by} +  \sum_{s=0}^{\infty} \vPhi_s\bL \vOmega_{t-s}^{\frac{1}{2}}\tilde{\mathbf{f}}_{t-s} + \sum_{s=0}^{\infty} \vPhi_s\bu_{t-s}^y.
\end{equation}

Since the latent factors act as structural shocks in our setup, in what follows we analyze how the system responds to unit shocks in $\tilde{\mathbf{f}}_{t}$. More specifically, we use the expression in~\eqref{eq:VMA} to derive structural impulse responses, forecast variance decomposition and historical decomposition. 

\subsection*{Structural Impulse Responses}

We first derive an expression for the response of $y_{i,t+l}$, the $i$-th element in $\by_{t+l}$, to unit shocks in $\tilde{\mathbf{f}}_t = (\tilde{f}_{1,t},\ldots, \tilde{f}_{r,t})'$ $l$ period ago:
\[
	\theta_{ij,l,t} \equiv \frac{\partial y_{i,t+l}}{\partial \tilde{f}_{j,t}}, 
\]
so that for each pair $(l,t)$, $\vTheta_{l,t} = (\theta_{ij,l,t})$ is of dimension $n\times r$. By differentiating \eqref{eq:VMA} with respect to $\tilde{\mathbf{f}}_t$, it is straightforward to see that the impulse response functions are given by:
\[
	\vTheta_{l,t}  = \vPhi_l \bL \vOmega_t^{\frac{1}{2}} = \left[\bA^{l}\right]_{1:n,1:n}\bL \vOmega_t^{\frac{1}{2}}.
\]
Since the variances of the latent factors are time-varying, these impulse response functions are technically also time-varying. However, the effect of the variance only scales the responses proportionally---the size of a unit shock changes over time. In practice we report only the impulse responses at a particular time $t$, e.g., $t=T$ with 
$\vTheta_{l,T}=\vPhi_l \bL \vOmega_T^{\frac{1}{2}}$.

\subsection*{Forecast Error Variance Decompositions}

Next, we develop an expression to account for the proportion of the forecast error variance or the mean squared prediction error (MSPE) that is due to the variation in the latent factors. To that end, let $\by_{t+l\gvn t}$ denote the optimal conditional forecast of $\by_{t+l}$ given the information up to time $t$. Then, using~\eqref{eq:VMA} it is follows that 
\[
	\by_{t+l}-\by_{t+l\gvn t}=\sum_{s=0}^{l-1} \vPhi_s \bL \vOmega_{t+l-s}^{\frac{1}{2}}\tilde{\mathbf{f}}_{t+l-s}+\sum_{s=0}^{l-1} \vPhi_s\bu_{t+l-s}^y.
\]
If we define $\vect{\Xi}_{l,t}=\vPhi_l\vSigma_t^{\frac{1}{2}}$, then the MSPE can be expressed as
\begin{align*}
	\text{MSPE}_t(l) & = \Em[(\by_{t+l}-\by_{t+l\gvn t})(\by_{t+l}-\by_{t+l\gvn t})'] \\
	& =\sum_{s=0}^{l-1} \vPhi_s \bL \vOmega_{t+l-s}^{\frac{1}{2}}\Em[\tilde{\mathbf{f}}_{t+l-s}\tilde{\mathbf{f}}_{t+l-s}'] \vOmega_{t+l-s}^{\frac{1}{2}}\bL'  \vPhi_s'+ \sum_{s=0}^{l-1} \vPhi_s\Em[\bu_{t+l-s}^y\bu_{t+l-s}^{y\prime}]\vPhi_s' \\
	& =\sum_{s=0}^{l-1} \vect{\Theta}_{s,t+l-s} \vect{\Theta}_{s,t+l-s}' +
	\sum_{s=0}^{l-1} \vect{\Xi}_{s,t+l-s} \vect{\Xi}_{s,t+l-s}'
	\equiv \text{MSPE}^{\bff}_t(l)+\text{MSPE}^{\bu}_t(l),
\end{align*}
where we have used the assumption that $\bu_{t}$ and $\tilde{\mathbf{f}}_{s}$ are mutually independent for all leads and lags. Hence, we have decomposed the MSPE into two components: one that can be attributed to the latent factors and the other to the idiosyncratic shocks. 

Since in our setup both the variances of the factors and the idiosyncratic shocks are time varying, the expression for MSPE depends on $t$. In practice, we focus on $t=T$ and compute 
$\text{MSPE}_T(l)$. Let $\theta_{ij,s,T+l-s}$ denote the $(i,j)$ element of $\vect{\Theta}_{s,T+l-s}$. Then, the contribution of the $j$-th factor to the MSPE of $y_{i,t}$, $i=1,\ldots,n$, at horizon $l$ is 
\[
	\text{MSPE}^{j,\bff}_{i,T}(l) =\sum_{s=0}^{l-1} \theta_{ij,s,T+l-s}^2. 
\]
Hence, we can further decompose the MSPE of $y_{i,t}$ attributed to the factors as
\[
	\text{MSPE}^{\bff}_{i,T}(l)  = \sum_{j=1}^r \text{MSPE}^{j,\bff}_{i,T}(l)=\sum_{j=1}^r \left(\sum_{s=0}^{l-1} \theta_{ij,s,T+l-s}^2 \right).
\]
It follows that the ratio $\text{MSPE}^{j,\bff}_{i,T}(l)/\text{MSPE}^{\bff}_{i,T}(l)$ measures the contribution of the $j$-th factor in forecasting the $i$-the variable at time $T$ $l$ periods ahead as a fraction of the MSPE attributed to the factors. 

\subsection*{Historical Decompositions}

Next, we develop expressions for historical decompositions. To that end, let $\dot{\by}_t$ denote the demeaned ${\by}_t$, i.e., $\dot{\by}_t = \by_t -\vmu_{\by}$. Then, it follows from~\eqref{eq:VMA} that one may approximate $\dot{\by}_t$ using
\[
	\hat{\dot{\by}}_t = \sum_{s=0}^{t-1} \vPhi_s \bL {\mathbf{f}}_{t-s} +
	\sum_{s=0}^{t-1}\vPhi_s \bu_{t-s}^y.
\]
For a covariance-stationary system, the approximation error becomes negligible for a sufficiently large $t$. To quantify how much the $j$-th factor explains the historically observed fluctuation in the $i$-th variable, let
\[
	\hat{\dot{y}}_{i,t}^{(j),\bff}=\sum_{s=0}^{t-1}\left(\mathbf{e}_{n,i}'\vPhi_s \bL\mathbf{e}_{r,j}\right) \left({\mathbf{e}_{r,j}'\mathbf{f}}_{t-s}\right), \quad 
	\hat{\dot{y}}_{i,t}^{\bu}=\sum_{s=0}^{t-1}\mathbf{e}_{n,i}'\vPhi_s \bu_{t-s}^y,
\]
where $\mathbf{e}_{m,k}$ denotes the $m\times 1$ vector with a 1 in the $k$-th coordinate and 0 elsewhere. Then, the historical decomposition of the $i$-th element of $\hat{\dot{\by}}_t$ can be expressed as
\[
  \hat{\dot{y}}_{i,t}=\sum_{j=1}^r \hat{\dot{y}}_{i,t}^{(j),\bff}+\hat{\dot{y}}_{i,t}^{\bu}.
\]
Hence, we have expressed $\hat{\dot{y}}_{i,t}$ as the summation of $r+1$ terms: the variations in $y_{i,t}$ that can be attributed to the $r$ factors and an additional `residual' term.

\newpage

\ifx\undefined\BySame
\newcommand{\BySame}{\leavevmode\rule[.5ex]{3em}{.5pt}\ }
\fi
\ifx\undefined\textsc
\newcommand{\textsc}[1]{{\sc #1}}
\newcommand{\emph}[1]{{\em #1\/}}
\let\tmpsmall\small
\renewcommand{\small}{\tmpsmall\sc}
\fi

\end{document}